\documentclass[12pt]{iopart}

\usepackage{amssymb}
\usepackage{tensor}
\usepackage{mathrsfs}
\usepackage{braket}
\usepackage{xfrac}
\usepackage{color}

\newcommand{\vc}[1]{\ensuremath{\mathbf{#1}}}
\newcommand{\pid}{\ensuremath{\mathrm{D}_t}}
\newcommand{\sd}{\ensuremath{\mathrm{D}}}
\newcommand{\nd}{\ensuremath{\mathscr{D}}}

\newcommand{\paren}[1]{\ensuremath{\left(#1\right)}}

\newcommand{\christ}{\ensuremath{\Gamma}}
\newcommand{\riemann}{\ensuremath{R}}
\newcommand{\ricci}{\ensuremath{R}}

\newcommand{\spricci}{\ensuremath{{}^{(d)}}R}
\newcommand{\espricci}{\ensuremath{{}^{(d)}}\tilde{R}}

\newcommand{\cschrist}{\ensuremath{\Gamma}}
\newcommand{\csriemann}{\ensuremath{\mathscr{R}}}
\newcommand{\csricci}{\ensuremath{\mathscr{R}}}
\newcommand{\csm}{\ensuremath{\mathscr{M}}}
\newcommand{\csp}{\mathscr{P}}

\newcommand{\dw}{\ensuremath{G}}

\newcommand{\lag}{\ensuremath{\mathscr{L}}}
\newcommand{\ham}{{\cal H}}
\newcommand{\lie}{\ensuremath{\mathcal{L}}}

\newcommand{\ep}{\lambda}
\newcommand{\p}{\varphi}
\newcommand{\sig}{\varepsilon}
\newcommand{\man}{\mathcal{M}}
\newcommand{\hyp}{\Sigma}

\newcommand{\rms}{\ensuremath{\mathrm{s}}}

\begin{document}
\title[Quantum gravitational corrections from the Wheeler-DeWitt equation\ldots]
{Quantum gravitational corrections from the Wheeler-DeWitt equation for scalar-tensor theories}
\author{Christian F Steinwachs$^1$ and Matthijs L van der Wild$^1$}
\address{$^1$ Physikalisches Institut, 
Albert-Ludwigs-Universit\"at Freiburg,
Hermann-Herder-Str. 3, 
79104,
Freiburg im Breisgau,
Germany}
\eads{\mailto{christian.steinwachs@physik.uni-freiburg.de}, \mailto{wild@physik.uni-freiburg.de}}

\begin{abstract}
We perform the canonical quantization of a general scalar-tensor theory and  
derive the first quantum gravitational corrections following from a semiclassical expansion of the Wheeler-DeWitt equation. The non-minimal coupling of the scalar field to gravity induces a derivative coupling between the scalar field and the gravitational degrees of freedom, which prevents a direct application of the expansion scheme. We address this technical difficulty by transforming the theory from the Jordan frame to the Einstein frame. We find that a large non-minimal coupling can have strong effects on the quantum gravitational correction terms. We briefly discuss these effects in the context of the specific model of Higgs inflation. 
\end{abstract}

\maketitle


\section{Introduction}\label{Sec:Intro}
Finding a consistent theory of quantum gravity is probably the most fundamental problem in theoretical physics.
There are many approaches to quantum gravity, see e.g. \cite{Kiefer:2007ria} for an overview. 
To a large extent, the difficulties in finding such a theory can be attributed to the absence of experimental guidance, 
as quantum gravitational effects are in general suppressed by inverse powers of the reduced Planck mass
\begin{equation}
M_{\mathrm{P}}=\sqrt{\frac{\hbar c}{8\pi G_{\mathrm{N}}}}\approx 10^{18}\,\frac{\text{GeV}}{c^2}\,.\label{eq.1.1}
\end{equation}
Therefore, quantum gravitational effects can be safely ignored for the elementary processes tested in 
terrestrial high-energy physics experiments and theoretically described by the Standard Model of particle physics.
Nevertheless, there are extreme physical situations, where quantum gravitational effects are expected to be 
become important or even dominant, such as the physics of black holes or the early universe. 

Any candidate theory of quantum gravity needs to satisfy at least two fundamental criteria. 
First, it must reproduce Einstein's classical theory of gravity in the semiclassical limit. 
Second, it must predict genuinely new quantum gravitational effects that are, at least in principle, measurable. 
Cosmology provides the natural testing ground for any theory of gravity. 
In particular, quantum gravitational effects produced during the inflationary stage, 
where the universe underwent a phase of accelerated expansion, 
might leave their imprint in the spectrum of the cosmic microwave background radiation.
In addition, the pessimistic conclusion that quantum gravitational effects are suppressed by inverse powers 
of the Planck mass \eref{eq.1.1} does not need to hold in models of modified gravity, such as scalar-tensor 
theories or $f(R)$-gravity, which are widely used in cosmology, see e.g. 
\cite{Sotiriou:2006hs,Clifton:2011jh,Nojiri:2017ncd,Nojiri:2010wj} and references therein.
In particular, in scalar-tensor theories with a strong non-minimal coupling \cite{Spokoiny:1984bd, Fakir:1990eg, Salopek:1988qh}, 
such as non-minimal Higgs inflation 
\cite{Bezrukov:2007ep, Barvinsky:2008ia,DeSimone:2008ei,Bezrukov:2008ej,Barvinsky:2009ii,Barvinsky:2009fy,Bezrukov:2010jz}, 
this might lead to an effective enhancement of quantum gravitational effects, 
which could bring them closer to observationally accessible scales. 

In the canonical approach to quantum gravity in $D=d+1$ dimensions, spacetime is foliated by a family of $d$-dimensional spatial hypersurfaces. The natural geometric phase space variables, following from the Arnowitt-Deser-Misner decomposition, are the $d$-dimensional metric field and its conjugated momentum, which both have a clear physical interpretation \cite{Arnowitt:1960es}. Promoting these variables to quantum operators results in the canonical theory of quantum gravity, called quantum geometrodynamics \cite{Wheeler:1957mu}.
The original $D$-dimensional diffeomorphism invariance of the gravitational action constrains the total Hamiltonian to vanish. This constraint on the total Hamiltonian can be divided into two parts: the momentum constraint which is the generator of the  $d$-dimensional spatial diffeomorphisms and the Hamiltonian constraint, which controls the dynamical evolution. Upon quantization in the Schr\"odinger representation, these classical constraints are turned into constraint operators. Implementing these constraints as suggested by Dirac \cite{Dirac1964}, the Hamiltonian constraint operator turns into to the Wheeler-DeWitt equation \cite{DeWitt:1967yk}, which determines the quantum dynamics of the system.

The canonical approach to quantum gravity comprises many problems -- both of conceptual and of technical nature.
To begin with, it is not even clear whether a Hilbert space structure exists for quantum gravity \cite{Kiefer:2007ria}. Connected to this are the problems associated to the definition of an inner product, which is required by the probabilistic interpretation of the quantum theory. 
Canonical quantum gravity is a non-perturbative approach, in which the Wheeler-DeWitt equation arises as an exact equation. The equation itself also comes with several technical problems ranging from factor ordering ambiguities to appearances of singular delta functions arising from functional derivatives evaluated at the same point. In most cases, the treatment of the full Wheeler-DeWitt equation therefore does not go beyond a formal level. To say the least, finding a general solution to the full Wheeler-DeWitt equation seems elusive.

Nevertheless, even if the whole approach is based on rather heuristic concepts and might suffer from a lack of mathematical rigor, there are still many interesting questions that can be addressed within the canonical formalism, such as for example the emergence of a semiclassical time from an otherwise timeless quantum world.
Moreover, aside from all the aforementioned problems, it can be expected that the Wheeler-DeWitt equation contains at least some valuable physical information, as it is possible to perform a systematic semiclassical expansion scheme, which reproduces the classical theory at the lowest orders. Therefore, higher order terms in such an expansion can be clearly attributed to quantum gravitational effects.
When applied to cosmology, these effects can even lead to measurable consequences, as has been investigated for example in \cite{Kiefer:2011cc, Bini:2013fea, Kamenshchik:2013msa, Kamenshchik:2014kpa, Brizuela:2015tzl, Brizuela:2016gnz}, where the impact of quantum gravitational corrections terms on the spectrum of the cosmic microwave background radiation has been investigated for a minimally coupled scalar field in a homogeneous and isotropic Friedmann-Lema\^\i tre-Robertson-Walker universe.
In this article we extend the quantum geometrodynamical formalism to a general scalar-tensor theory and focus on the semiclassical expansion of the Wheeler-DeWitt equation. We derive the first quantum gravitational corrections and investigate how they are affected by a non-minimal coupling.

The paper is structured as follows.
In \Sref{Sec:ADM}, we perform the $d+1$ decomposition for a general scalar-tensor theory with a non-minimal coupling in the Jordan frame parametrization.
In \Sref{Sec:CanonicalFormulationHamilton}, we formulate the classical theory in the Hamiltonian framework and derive the momentum and Hamiltonian constraints.
In \Sref{Sec:QuantumHamiltonian}, we perform the transition to the quantum theory and derive the Wheeler-DeWitt equation.
In \Sref{Sec:WeightingEFJF}, we express the Wheeler-DeWitt equation in the Einstein frame parametrization, where it becomes diagonal, and perform the weighting required for the application of the Born-Oppenheimer approximation scheme. Finally, we re-express the weighted Wheeler-DeWitt operator in the Jordan frame parametrization.
In \Sref{Sec:SemiClassicalExpansion}, we perform the semiclassical approximation up to the order where the first quantum gravitational corrections arise and discuss the impact of the non-minimal coupling. 
In \Sref{Sec:Conclusions}, we summarize our results and briefly discuss possible applications.  

Technical details are provided in several appendices.
In \ref{App:GeometryConfigurationSpace}, we present detailed expressions for objects related to the geometry of configuration space.   
In \ref{App:ConformalTransformation}, we provide the explicit transformation laws resulting from a conformal transformation of the metric field.
In \ref{App:JFtoEF}, we present the details of the transition from the Jordan frame to the Einstein frame.

\section{Scalar-tensor theory and foliation of spacetime}\label{Sec:ADM}
The action functional of a general scalar-tensor theory in $D=d+1$ dimensions with a scalar field $\p$ non-minimally coupled to gravity, can be parametrized in terms 
of three arbitrary functions $U(\p)$, $G(\p)$ and $V(\p)$, 
\begin{equation}
\label{eq.2.1}
S[g,\p]=\int_{\man}\rmd^Dx\,\left(\sig g\right)^{\sfrac{1}{2}}
\left(U\,R-\frac{1}{2} G\,\nabla_\mu\p\nabla^\mu\p-V\right)\,.
\end{equation}
Here, $U$ is the non-minimal coupling, $G$ parametrizes 
a non-canonically normalized kinetic term and $V$ is the 
scalar field potential. We assume that the manifold $\man$ is 
globally hyperbolic and endowed with the $D$-dimensional metric 
$g_{\mu\nu}$ and a metric compatible affine connection $\nabla_{\mu}$. 
The signature of $\man$ is determined by the constant parameter $\sig=\pm1$, 
where plus corresponds to Euclidean signature and minus to
Lorentzian signature. The Riemannian curvature is defined by
\begin{equation}
{}^{(D)}\tensor{\riemann}{^{\rho}_\sigma_\mu_\nu}v^{\sigma}
=[\nabla_{\mu},\,\nabla_{\nu}]v^{\rho},\qquad v \in T\man.
\end{equation}
A point $X\in\man$ can be described by local coordinates $X^{\mu}$. 
In order to express the action \eref{eq.2.1} in the Hamiltonian formalism, 
we foliate the $D$-dimensional ambient space $\man$ by a one-parameter 
family of $d$-dimensional hypersurfaces $\hyp_t$ of constant time $t$. 
Thus, the hypersurfaces $\hyp_t$ are the level surfaces of a globally 
defined smooth scalar time field $t$. 
The gradient of $t$ defines a natural unit covector field
\begin{equation}
n_\mu:=\sig\frac{\nabla_\mu t}{\sqrt{\sig g^{\mu\nu}\nabla_\mu t\nabla_\nu t}},
\qquad 
g^{\mu\nu}n_\mu n_\nu=\sig\,.
\end{equation}
At each point in $\hyp_t$, the normal vector field
\begin{equation}
n^{\mu}=g^{\mu\nu}n_{\nu}\,,
\end{equation}
is orthogonal to $\hyp_t$ and allows an orthogonal decomposition of tensor fields with respect to $n^{\mu}$. In particular, the ambient metric decomposes as 
\begin{equation}
g_{\mu\nu}=\sig n_\mu n_\nu+\gamma_{\mu\nu}\,.
\end{equation}
Here, $\gamma_{\mu\nu}$ is the tangential part of $g_{\mu\nu}$, 
that is $\gamma_{\mu\nu}n^{\mu}=0$. The hypersurfaces $\hyp_t$
can be considered as the embeddings of an intrinsically 
$d$-dimensional manifold $\widehat{\hyp}_t$ into the ambient 
space $\man$. A point $x\in\widehat{\hyp}$ can be described 
by local coordinates $x^a$. Thus, we can parametrize the $D$-dimensional 
coordinate $X^{\mu}=X^{\mu}(t,x^{a})$ in terms of the time field $t$ and 
the $d$-dimensional coordinates $x^{a}$. The change of $X^{\mu}$ with 
respect to $t$ and $x^{a}$ can be described by the coordinate one-form 
\begin{equation}
\rmd X^\mu=\frac{\partial X^{\mu}(t,x)}{\partial t}\rmd t
+\frac{\partial X^{\mu}(t,x)}{\partial x^{a}}\rmd x^{a}
=t^{\mu}\rmd t+\tensor{e}{^{\mu}_{a}}\rmd x^{a},\label{eq.2.2}
\end{equation}
where we have defined the time vector field $t^{\mu}$ 
and the soldering form $\tensor{e}{^{\mu}_{a}}$ as
\begin{equation}
t^{\mu}:=\frac{\partial X^\mu(t,x^{i})}{\partial t}:=N\,n^{\mu}+N^{\mu},
\qquad 
\tensor{e}{_{a}^{\mu}}:=\frac{\partial X^\mu(t,x)}{\partial x^a}.
\label{eq.2.3}
\end{equation}
Under the orthogonal decomposition, the time vector field $t^{\mu}$
has a component $N=N(t,x)$ in the normal direction $n^{\mu}$, which is called 
``lapse function'' and a component $N^{\mu}=N^{\mu}(t,x)$ tangential to $\hyp_t$, which is called ``shift vector''. 
The soldering form $\tensor{e}{^{\mu}_{a}}$ can be thought of as a tangential vector 
with respect to the $\mu$ index, that is $\tensor{e}{^{\mu}_{a}}n_{\mu}=0$, and a $d$-dimensional
vector with respect to the $a$ index. It can  be used to pull back tangential
tensors in $\man$ to tensors in $\widehat{\hyp}_t$. In particular, we have
\begin{eqnarray}
\gamma_{ab}&:=\tensor{e}{^{\mu}_{a}}\tensor{e}{^{\nu}_b}\gamma_{\mu\nu},\qquad 
&N^{a}:=\tensor{e}{_{\mu}^{a}}N^{\mu},\\
\delta^{\mu}_{\nu}&=\tensor{e}{^{\mu}_{a}}\tensor{e}{_{\nu}^{a}},       \qquad 
&\delta^{a}_{b}=\tensor{e}{^{a}_{\mu}}\tensor{e}{^{\mu}_{b}}\,.
\end{eqnarray}
The ambient space coordinate one-form \eref{eq.2.2} 
can therefore be written as 
\begin{equation}
\rmd X^\mu=Nn^\mu\rmd t+\tensor{e}{_{a}^{\mu}}\left(N^a\rmd t+\rmd x^a\right)\,.
\end{equation}
The ambient space line element then acquires the Arnowitt-Deser-Misner form
\begin{eqnarray}
\rmd s^2&=g_{\mu\nu}\rmd X^{\mu}\rmd X^{\mu}\nonumber\\
&=\left(\sig N^2+\gamma_{ab}N^aN^b\right)\rmd t^2
+2\gamma_{ab}N^a\rmd x^b\rmd t+\gamma_{ab}\rmd x^a\rmd x^b\,.
\end{eqnarray} 
The volume element is found to decompose as
\begin{equation}
\left(\sig g\right)^{\sfrac{1}{2}}=N\gamma^{\sfrac{1}{2}}\,.\label{eq.2.3.6}
\end{equation}
On $\widehat{\hyp}_t$, the affine connection $D_a$ compatible 
with the metric $\gamma_{bc}$ defines the $d$-dimensional curvature
\begin{equation}
{}^{(d)}\tensor{R}{^a_b_c_d}v^{b}=[\sd_c,\,\sd_d]v^{a}\,,\qquad v\in T\widehat{\hyp}_t\,.
\end{equation} 
The relation between the $D$-dimensional Ricci scalar ${}^{(D)}R$ and 
the $d$-dimensional Ricci scalar ${}^{(d)}R$ is given by the Gauss-Codazzi
equation, see e.g. \cite{Kuchar:1976b},
\begin{equation}
{}^{(D)}R=\spricci+\sig\paren{K^2-K_{ab}K^{ab}}+2\sig\left(\sd_b-\sig a_b\right)a^b-2\sig\left(\pid+K\right)K\,.\label{eq.2.4}
\end{equation}
Here, $\spricci$ is the intrinsic $d$-dimensional curvature, which is 
calculated using the induced metric $\gamma_{ab}$,
while $K_{ab}$ is the extrinsic curvature and $K$ its trace
\begin{equation}
K_{ab} :=\frac{1}{2N}\left[\partial_t\gamma_{ab}-\paren{\lie_{\vc{N}}\gamma}_{ab}\right]
=\frac{1}{2}D_t\gamma_{ab}\,,\qquad K:=\gamma^{ab}K_{ab}\,.
\end{equation}
We have introduced the covariant reparametrization invariant time derivative
\begin{equation}
\label{eq.2.5}
\pid:=\frac{1}{N}\paren{\partial_t-\lie_{\vc{N}}}\,,
\end{equation}
where $\lie_{\vc{N}}$ is the Lie derivative along the spatial
shift vector $\vc{N}=N^a\partial_a$.
Finally, the decomposition \eref{eq.2.4} involves the acceleration vector $a_b$, which is defined as
\begin{equation}
a_b:=-\sig\sd_b\log{N}\,.\label{eq.2.6}
\end{equation}
Using \eref{eq.2.3.6}-\eref{eq.2.6}, the action \eref{eq.2.1} can be expressed in terms of intrinsic $d$-dimensional tensors,
\begin{equation}
S[N,\mathbf{N},\gamma,\p]=\int\rmd t\,L=\int \rmd t \rmd^dx\,\lag\,.\label{eq.2.7}
\end{equation}
Here, $L$ and $\lag$ are the Lagrangian and the Lagrangian density.
Up to boundary terms, the Lagrangian density for \eref{eq.2.1} is explicitly given by
\begin{eqnarray}
\fl
\lag=-N\gamma^{\sfrac{1}{2}}\Big[\sig U\dw^{abcd}K_{ab}K_{cd}-U\,\spricci-2\sig U_{1}K\pid\p-2\Delta U\nonumber\\
 +\frac{G}{2}\left(\sig\pid\p\pid\p+\sd_a\p\sd^a\p\right)+V\Big]\,.
\end{eqnarray}
Here, $\Delta:=-\gamma^{ab}\sd_a\sd_b$ denotes the $d$-dimensional Laplacian 
and we have introduced the ``dedensitized'' DeWitt metric $G^{abcd}$ and its inverse $G_{abcd}$,
\begin{equation}
\dw^{abcd}:=\gamma^{a(c}\gamma^{d)b}-\gamma^{ab}\gamma^{cd}\,,\qquad\dw_{abcd}=\gamma_{a(c}\gamma_{d)b}-\frac{1}{d-1}\gamma_{ab}\gamma_{cd}\,, 
\end{equation}
which satisfy
\begin{equation}
\dw_{abkl}\dw^{klcd}=\delta_{ab}^{cd}:=\delta_{(a}^c\delta_{b)}^d:=\frac{1}{2}\left(\delta_{a}^c\delta_{b}^d+\delta_{b}^c\delta_{a}^d\right)\,.
\end{equation}
In addition, we denote derivatives of a function $f(\p)$ with respect to its argument by
\begin{equation}
f_n(\varphi):=\frac{\partial^n f(\varphi)}{\partial \varphi^n}\,.
\end{equation}

\section{Canonical formulation and Hamiltonian constraint}\label{Sec:CanonicalFormulationHamilton}
We introduce a compact notation for the dynamical configuration space variables\footnote{Note that the lapse function and the shift vector are not dynamical degrees of freedom.} $q^A$ and their velocities $\partial_tq^A$, where the superindex $A$ labels the corresponding components
\begin{equation}
\left(q^{A}\right)=
\left(
\begin{array}{c}
\gamma_{ab}\\
\p
\end{array}
\right),\qquad
\left(\partial_tq^{A}\right)=
\left(
\begin{array}{c}
\partial_t\gamma_{ab}\\
\partial_t\p
\end{array}
\right).
\end{equation}
In this compact notation the Lagrangian density in \eref{eq.2.7} takes the form
\begin{equation}
 \lag=\frac{1}{2}\partial_tq^A\csm_{AB}\,\partial_tq^B+\ldots,
 \end{equation}
where the configuration space metric $\csm_{AB}$ can be read off from the terms quadratic in the velocities and the dots indicate lower order time derivatives terms. In components $\csm_{AB}$ has the block matrix structure
\begin{equation}
\label{eq.3.1}
\left(\csm_{AB}\right)=\frac{\sig\gamma^{\sfrac{1}{2}}}{N}
\left(
\begin{array}{cc}
-\frac{U}{2}G^{abcd} & U_{1}\gamma^{ab}\\
U_{1}\gamma^{cd} & -G
\end{array}
\right)\,.
\end{equation}
Note the somewhat unorthodox inclusion of the inverse lapse function into the definition of the configuration space metric \eref{eq.3.1}. In principle, time reparametrization invariance suggests to associate with each factor of time $t$ a factor of the lapse function such as the inverse powers of $N$ in the covariant time derivative \eref{eq.2.5}. Similarly, one would associate a factor of $N$ with the time differential $\rmd t$ in \eref{eq.2.7}. The inclusion of the lapse function in \eref{eq.3.1} will become clear in \Sref{Sec:Einstein frame}, where we discuss the transition between two particular parametrizations of the fields. 
In general, we consider the configuration space formally as a differentiable manifold and provide 
a list of the associated geometrical objects in \ref{App:GeometryConfigurationSpace}.

In terms of the covariant time derivative \eref{eq.2.5}, the Lagrangian density \eref{eq.2.7} acquires the compact form
\begin{equation}
\label{eq.3.2}
\lag=\frac{N^2}{2}\,\pid q^A\csm_{AB}\,\pid q^B-\csp\,,
\end{equation}
where $\pid$ acts componentwise on the $q^{A}$.\footnote{The tensor $\gamma_{ab}$, when 
viewed as metric field is defined with covariant spatial indices, despite 
its contravariant superindex when viewed as a configuration space coordinate.}  
The potential $\csp$ is defined as
\begin{eqnarray}
\csp:=&\csp_{\gamma}+\csp_{\p}\,,\\
\csp_{\gamma}:=&N\,P_{\gamma}:=-N\gamma^{\sfrac{1}{2}}U\left[\spricci+2\frac{\Delta U}{U}+\frac{d}{d-1}\frac{\sd_{a}U\sd^{a}U}{U^2}\right]\,,\\
\csp_{\p}:=&N\,P_{\p}:=N\gamma^{\sfrac{1}{2}}\left[\frac{1}{2 s }\sd_{a}\p\sd^{a}\p+V\right]\,,
\end{eqnarray}
where we have introduced the abbreviation
\begin{equation}
\label{eq.3.3}
 s:=\frac{U}{GU+\frac{2d}{d-1}U_{1}^2}\,.
\end{equation}
The momenta can be calculated directly from \eref{eq.3.2},
\begin{eqnarray}
p_A&=\frac{\partial \lag}{\partial(\partial_tq^A)}=N\csm_{AB}\pid q^B.\label{eq.3.4}
\end{eqnarray}
In components, the momenta read
\begin{eqnarray}
\left(p_A\right)&=\left(
\begin{array}{c}
p^{ab}_{\gamma}\\
p_{\p}
\end{array}\right)
=
\sig\gamma^{\sfrac{1}{2}}
\left(
\begin{array}{c}
-U\dw^{abcd}K_{cd}+U_{1}\gamma^{ab}\pid\p\\
2U_{1}\gamma^{ab}K_{ab}-G\pid\p
\end{array}
\right)\,.
\end{eqnarray}
We can invert relation \eref{eq.3.4} and obtain
\begin{equation}
\pid q^{A}=N^{-1}\csm^{AB}p_B\,,
\end{equation}
where the inverse of the configuration space metric $\csm^{AB}$ is defined by
\begin{equation}
\csm_{AC}\csm^{CB}=\delta_{A}^{B},\qquad \left(\delta_{A}^{B}\right)=
\left(
\begin{array}{cc}
\delta_{ab}^{cd} & 0\\
0                & 1
\end{array}
\right)\,.
\end{equation}
Here, the $(\delta^{A}_{B})$ denote the components of the identity matrix on the configuration space.\footnote{For the inverse $\csm^{AB}$ to exist, it is required that $d>1$, $U\neq0$, $\det\gamma\neq0$ and $G\neq-2d/(d-1)U_{1}^2/U$.}
The components of the inverse configuration space metric $\csm^{AB}$ are given explicitly by
\begin{equation}
\label{eq.3.1inv}
\left(\csm^{AB}\right)=\sig N\gamma^{-\sfrac{1}{2}}
\left(
\begin{array}{cc}
-\frac{2}{U}\dw_{abcd}- \frac{4\, s  }{(d-1)^2}\left(\frac{U_{1}}{U}\right)^2\gamma_{ab}\gamma_{cd} & \frac{2 s  }{d-1}\frac{U_{1}}{U}\gamma_{ab}\\
\frac{2 s }{d-1}\frac{ U_{1}}{U}\gamma_{cd} & - s 
\end{array}
\right).
\end{equation}
The Hamiltonian density $\ham$ is obtained by the Legendre transformation of \eref{eq.3.2},
\begin{eqnarray}
\ham&=p_A\partial_tq^A-\lag=p_A\left(N\,\pid q^A+\lie_{\vc{N}} q^A\right)-\lag\nonumber\\
&=\frac{1}{2}p_A\csm^{AB}p_B+\csp+p_A\lie_{\vc{N}}q^A\,.\label{eq.3.5}
\end{eqnarray}
The total Hamiltonian is given by the spatial integral of \eref{eq.3.5},
\begin{equation}
\label{eq.3.6}
H:=\int\rmd^dx\,\ham:=\int\rmd^dx\paren{N\ham_\perp+N^a\ham_a}\,,
\end{equation}
As a consequence of the $D$-dimensional diffeomorphism invariance of \eref{eq.2.1}, the total Hamiltonian \eref{eq.3.6} is constrained to vanish. The constraint character becomes manifest in the last equality of \eref{eq.3.6}, where we have written $\ham$ as the sum of the Hamiltonian constraint $\ham_\perp$ and the momentum constraint $\ham_a$ together with the lapse function $N$ and shift vector $N^{a}$, which act as Lagrange multipliers. Explicitly, the constraints are given by 
\begin{eqnarray}
\ham_\perp&=-\frac{\sig}{U\sqrt{\gamma}}\dw_{abcd}\,p^{ab}_{\gamma}p^{cd}_{\gamma}+P_{\gamma}
-\frac{\sig}{2}\frac{ s }{\sqrt{\gamma}}\left(p_{\p}-\frac{2}{d-1}\frac{U_{1}}{U}p_{\gamma}\right)^2+ P_{\p}\,,
\label{eq.3.7}\\
\ham_a&=-2\gamma_{a(b}\sd_{c)}p^{bc}_{\gamma}+p_\p\sd_a\p\,.\label{eq.3.8}
\end{eqnarray}
Note that we have defined the trace of the gravitational momentum $p_{\gamma}=\gamma_{ab}\,p_{\gamma}^{ab}$. 
The momentum constraint $\ham_{a}$ is the generator of $d$-dimensional diffeomorphisms, while the dynamical evolution is controlled by the Hamiltonian constraint $\ham_{\perp}$. The expressions \eref{eq.3.7} and \eref{eq.3.8} coincide with those obtained in \cite{Weenink:2010rr}.

\section{Quantum Theory and Wheeler-DeWitt equation}\label{Sec:QuantumHamiltonian}
As mentioned in the introduction, it is unclear whether a Hilbert space structure is a prerequisite for canonical quantum gravity. Nevertheless, we assume that there is at least an auxiliary underlying Hilbert space with Schr\"odinger type inner product
\begin{equation}
\braket{\Phi|\Psi}:=\int\rmd q\,\csm^{1/2}\,\bar{\Phi}[q]\,\Psi[q]\,.\label{eq.4.1}
\end{equation}
Here, the wave functional $\Psi[q]=\braket{q|\Psi}$ corresponds to the 
Schr\"odinger representation of the state $\ket{\Psi}$ and $\csm$ is the determinant of the configuration space metric $\csm_{AB}$. Note that the naive definition \eref{eq.4.1} involves the integration over all configurations $q^{A}$, including the unphysical ones, as \eref{eq.4.1} is not an inner product on the space of the solutions to the constraints \cite{Kiefer:2007ria}. 
Related to the inner product is the question of unitarity.
This is a complicated problem in the context of quantum gravity and can be discussed at various levels \cite{Barvinsky:1993jf}.
In particular, if quantum theory is a universal concept, a probabilistic interpretation would require a unitary evolution at the most fundamental level, including the gravitational degrees of freedom.
However, in this article we only focus on the semiclassical approximation and therefore do not address these fundamental questions. 

In the quantum theory, the conjugated phase space variables $q^{A}$ and $p_{B}$
are promoted to operators $\hat{q}^{A}$ and $\hat{p}_B$, which satisfy the canonical commutator relations
\begin{equation}
\left[\hat{q}^{A},\,\hat{p}_{B}\right]=
\rmi\,\delta^{A}_{B}\,,\qquad \left[\hat{q}^{A},\hat{q}^B\right]=\left[\hat{p}_{A},\hat{p}_B\right]=0\,,\label{eq.4.2}
\end{equation}
In the position space representation, the position operator $\hat{q}^{A}$ acts multiplicatively, while the momentum operator $\hat{p}_B$ acts as a derivative operator
\begin{equation}
\hat{p}_{A}:=-\rmi\,\csm^{-1/4}\frac{\delta}{\delta q^{A}}\csm^{1/4}\,.
\end{equation}
This representation of the momentum operator is formally self-adjoint with respect to the inner product \eref{eq.4.1} and satisfies the canonical commutation relation \eref{eq.4.2}, see \cite{DeWitt:1952}.
The operator versions of the classical constraints \eref{eq.3.7} are defined by replacing the classical phase space variables by their quantum operators 
\begin{equation}
\hat{\ham}_{\perp}:=\ham_{\perp}(\hat{q},\hat{p})\,,\qquad \hat{\ham}_{a}:=\ham_{a}(\hat{q},\hat{p})\,.
\end{equation}
This procedure is ambiguous due to factor ordering problems, which arise because of \eref{eq.4.2}. 
In particular, for the transition from the classical Hamiltonian constraint
\begin{equation}
 \ham_\perp=\frac{1}{N}\left(\frac{1}{2}p_A\csm^{AB}p_B+\csp\right)\label{eq.4.4}
\end{equation}
to the quantum Hamiltonian constraint, this factor ordering ambiguity can be traced back to the non-commutativity of the configuration space metric with the momentum operator
\begin{equation}
[\csm_{AB}(\hat{q}),\hat{p}_C]\neq0\,.
\end{equation}
The factor ordering ambiguity does not affect the principal part of the Hamiltonian constraint operator $\hat{\ham}_{\perp}$ -- only its lower derivative terms. It can be partially addressed by adopting the covariant Laplace-Beltrami factor ordering, which effectively corresponds to replacing the quadratic form $p_{A}\csm^{AB}(q)p_{B}$ in \eref{eq.4.4} by the symmetric combination\footnote{Instead of $\square$, it is also possible to consider a generalized Laplacian with a potential term. This includes in particular the conformal Laplacian, where the potential is proportional to the configuration space Ricci scalar, derived in \eref{csriccis}.} 
\begin{eqnarray}
\csm^{-1/4}\hat{p}_A\csm^{1/4}\,\csm^{AB}\,\csm^{1/4}\hat{p}_{B}\csm^{-1/4}
&=-\csm^{AB}\nabla_{A}\nabla_{B}=:-\square\,.\label{eq.4.5}
\end{eqnarray}
Here, we have abbreviated ordinary functional derivatives by $\delta_{A}$ and introduced the covariant functional derivatives $\nabla_{A}$, defined with respect to the Christoffel connection of the configuration space metric,
\begin{eqnarray}
\delta_{A}\Psi=\frac{\delta\Psi}{\delta q^{A}},\qquad \tensor{\cschrist}{^C_{AB}}=\frac{1}{2}\csm^{CD}\paren{\delta_A\csm_{DB}+\delta_B\csm_{AD}-\delta_D\csm_{AB}}\,.
\end{eqnarray}
The Hamiltonian constraint operator with the factor ordering \eref{eq.4.5} can be written compactly as
\begin{equation}
\label{eq.4.6}
 \hat{\ham}_\perp=\frac{1}{N}\left(-\frac{1}{2}\square+\csp\right)\,.
\end{equation}
In the following, we refer to \eref{eq.4.6} as the Wheeler-DeWitt operator.
Using the geometrical quantities provided in \eref{ccsm1}-\eref{ccsm6}, we obtain the explicit form of the Laplace-Beltrami operator \eref{eq.4.4} and hence for the Wheeler-DeWitt operator \eref{eq.4.6},
\begin{eqnarray}
\fl
\hat{\ham}_{\perp}=\frac{\sig}{\sqrt{\gamma}U}G_{abcd}\frac{\delta^2}{\delta\gamma_{ab}\delta\gamma_{cd}}+\frac{\sig}{\sqrt{\gamma}U}\frac{(3d+5)(d-2)}{8(d-1)}\gamma_{ab}\frac{\delta}{\delta\gamma_{ab}}+P_{\gamma}\nonumber\\
+\frac{\sig}{2}\frac{ s }{\sqrt{\gamma}}\nd^2+\frac{\sig}{4}\frac{ s }{\sqrt{\gamma}}\left(\frac{s_{1}}{s}-\frac{d}{2}\frac{U_{1}}{U}\right)\nd+P_{\p}\,.\label{eq.4.7}
\end{eqnarray}
Here, we have introduced the combined derivative
\begin{equation}
 \nd:=\frac{\delta}{\delta\p}-\frac{2}{d-1}\frac{U_{1}}{U}\gamma_{ab}\frac{\delta}{\delta \gamma_{ab}}\,.\label{eq.4.8}
\end{equation}
Note, that since the Laplace-Beltrami operator $\square$ and the potential $\csp$ are both proportional to $N$, the explicit form of $\hat{\ham}_{\perp}$ is independent of the lapse function.

We follow the quantization prescription for constrained systems, 
proposed by Dirac \cite{Dirac1964}, where the quantum constraints are implemented by demanding that physical 
states are annihilated by the quantum constraint operators.
The implementation of the momentum constraint operator $\hat{\ham}_{a}$ ensures that the wave functional $\ket{\Psi}$ is invariant under $d$-dimensional diffeomorphisms
\begin{equation}
\hat{\ham}_{a}\ket{\Psi}=0\,.\label{eq.4.9}
\end{equation}
The configuration space modulo the $d$-dimensional diffeomorphisms is called ``superspace'' \cite{Wheeler1968,Dewitt:1969ke}.
The implementation of the Hamiltonian constraint operator $\hat{\ham}_{\perp}$, which governs the quantum dynamics of the wave functional $\Psi$, leads to the Wheeler-DeWitt equation
\begin{equation}
 \hat{\ham}_{\perp}\ket{\Psi}=0\,.\label{eq.4.10}
\end{equation}
Finally, let us comment on factors of $\delta^{(d)}(0)$, which we have suppressed in the considerations so far. We assume the spatial $d$-dimensional delta function $\delta^{(d)}(x_A,x_B)$ to be a scalar bidensity with zero  weight at the first argument and unit weight at the second argument, where $x_{A}$ and $x_{B}$ are the corresponding spatial coordinates. Given the fundamental identity
\begin{equation}
\frac{\delta q^{A}(x_A)}{\delta q^{B}(x_B)}=\delta^{A}_{B}\,\delta^{(d)}(x_A,x_B)\,,
\end{equation}
whenever a functional derivative acts on a local background quantity at the same point $x_A=x_B$, this leads to a factor of $\delta^{(d)}(0)$ \cite{DeWitt:1967yk}.
The singular factors of undifferentiated delta functions at the same point have to be regularized.
In \cite{DeWitt:1967yk}, it was suggested to adopt a regularization scheme where field operators at the same point can be freely commuted, which effectively corresponds to $\delta^{(d)}(0)=0$. Practically, for the Wheeler-DeWitt operator \eref{eq.4.7}, this means that the kinetic part is reduced to its principal (highest derivative) part, where all functional derivatives only act on the wave functional -- not on local background coefficients.
We do not adopt any regularization scheme at this point. Instead, we carry all factors of $\delta^{(d)}(0)$ through the calculation but suppress their explicit occurrence for notational reasons. At each step, explicit factors $\delta^{(d)}(0)$ can be restored easily by dimensional considerations.

\section{Weighting and transition between Jordan frame and Einstein frame}\label{Sec:WeightingEFJF}
In general, it is difficult to find an exact solution $\Psi[\gamma,\p]$ to the Wheeler-DeWitt equation \eref{eq.4.10}.
Moreover, the quantum theory obtained by the naive definition of the inner product and the Dirac quantization scheme is not complete \cite{Barvinsky:1993jf,Barvinsky:2013aya}. In addition, an exact solution to the Wheeler-DeWitt equation requires suitable boundary conditions. These boundary conditions are the main subject of quantum cosmology. Among many choices the most prominent and physically best motivated proposals seem to be the no-boundary and tunneling conditions, see \cite{Hawking:1982cz,Hartle:1983ai,Hawking:1983hj,Linde:1983mx,Rubakov:1984ki,Zeldovich:1984vk,Vilenkin:1984wp,Vilenkin:1987kf}.
Given a sensible definition of a probability measure, one might even extract predictions and consistency equations from quantum cosmology, see e.g. \cite{Barvinsky:1994hx, Vilenkin:1998rp,Barvinsky:2009jd,Calcagni:2014xca}. 

In this article, we are not concerned with finding an exact solution to the full non-perturbative Wheeler-DeWitt equation.
Instead, we aim to construct the semiclassical branch of the wave functional $\Psi$, which is limited to a restricted region in configuration space.
We use a combined Born-Oppenheimer/WBK approximation in order to perform a systematic expansion of the full Wheeler-DeWitt equation \cite{Gerlach:1969ph, Lapchinsky:1979fd, Brout:1987ya, Brout:1988ku, Banks:1984cw, Kiefer:1991,Bertoni:1996ms, Kamenshchik:2017kfs} which allows to extract the first quantum gravitational correction terms \cite{Kiefer:1991}.

\subsection{Born-Oppenheimer approximation}
The Born-Oppenheimer approximation is well known in quantum mechanical multi-particle 
systems and based on a distinction between ``heavy'' degrees of freedom $Q$ with mass $m_{Q}$ and ``light'' degrees of freedom $q$ with mass $m_{q}$ \cite{BornOppenheimer1927}.
The difference in the characteristic mass scales, expressed in terms of the dimensionless parameter
\begin{equation}
\ep:=\frac{m_{Q}}{m_{q}}\gg 1\,,\label{eq.5.2}
\end{equation}
implies that the heavy and light degrees of freedom vary on different characteristic time scales and might therefore be interchangeably associated with the ``slow'' and ``fast'' degrees of freedom. 
In a more abstract context, $\ep$ represents a formal parameter which can be used to implement the distinction between background (slow) and fluctuation (fast) degrees of freedom and which can be set to one after the expansion has been performed. 
The distinction between slow and fast variables allows to make a product ansatz for the total wave function
\begin{equation}
\Psi(Q,q)=\chi(Q)\,\psi(q;Q).
\end{equation}
Here $\chi(Q)$ is the wave function for the slow degrees of freedom $Q$, for which a subsequent WKB approximation is performed. In contrast, the wave function  $\psi(q;Q)$ for the fast degrees of freedom $q$ is treated as fully quantum and only depends parametrically on the $Q$ variables.

Practically, the semiclassical expansion can be systematically performed by the following operations. First, the distinction between heavy and light degrees of freedom can be implemented by assigning different relative weight factors for the individual terms in the Hamilton operator by rescaling each factor of $m_{Q}$ by a power of $\ep$,
\begin{equation}
H(q,Q)\to H_{\ep}(q,Q)\,,\label{eq.5.3}
\end{equation}
where the weighted Hamiltonian $H_{\ep}$ has the schematic structure
\begin{equation}
H_{\ep}(q,Q)=\frac{1}{\lambda}\frac{P_Q^2}{2 m_Q}+V(Q)+\frac{p_q^2}{2m_q}+W(q,Q)\,.
\end{equation}
Here $P_{Q}$ and $V(Q)$ are the momentum and self-interaction potential of the heavy variables $Q$, while $p_q$ and $W(q,Q)$ are the momentum and potential of the fast variables $q$. The latter includes the self-interaction among the $q$'s as well as the interaction between the $q$'s and the $Q$'s.
Second, in addition to the weighting of the Hamilton operator \eref{eq.5.3}, we make an ansatz for the wave function $\Psi(q,Q)$ in the form of a formal power series in $\ep$,
\begin{eqnarray}
\label{eq.5.4}
\Psi(q,Q)&=\exp\rmi\left[\ep S_0(q,Q)+S_1(q,Q)+\ep^{-1}S_2(q,Q)+\ldots\right]\,.
\end{eqnarray}

In the quantum gravitational context, the semiclassical expansion can be obtained by inserting the weighted Hamilton operator $\hat{H}_{\ep}$ together with the ansatz \eref{eq.5.4} into the Wheeler-DeWitt equation \eref{eq.4.10}. Collecting terms of equal power in $\ep$ and demanding that they vanish separately, leads to a sequence of equations for $S_0$, $S_1$, $S_2$, etc. Solving these equations consecutively order by order, the wave functional $\Psi$ can be reconstructed to the accuracy given by the respective order in $\ep$. 
For the system of a scalar field $\p$, minimally coupled to gravity in $D=d+1$ dimensions the slow and fast degrees of freedom $Q$ and $q$ are then associated with the spatial metric $\gamma_{ab}$ and the scalar field $\p$, respectively.
The weighting procedure then corresponds to the association of the expansion parameter $\ep$ with each occurence the squared Planck mass $U_0=M_{\mathrm{P}}^2/2$.

At the highest order of the expansion ${\cal O}(\ep^2)$, one finds that $S_0(\gamma)$ is a function of $\gamma_{ab}$ only. This is consistent with the association of the degrees of freedom in $\gamma_{ab}$ as the slow variable.
At the next order ${\cal O}(\ep)$, one obtains an equation for $S_0(\gamma)$. Since the equation is of the Hamilton-Jacobi type, one recovers in a natural way the notion of a semiclassical time from the timeless Wheeler-DeWitt equation.
At order ${\cal O}(\ep^{0})$ one obtains an equation for $S_1(\gamma,\p)$, which can be formulated as a Schr\"odinger equation for the light scalar field degree of freedom $\p$, where the time parameter $t$ is identified with the semiclassical time and is effectively provided by the slowly changing background geometry, see e.g. \cite{Gerlach:1969ph, Lapchinsky:1979fd, Brout:1987ya, Brout:1988ku, Banks:1984cw, Kiefer:1991,Bertoni:1996ms,Kamenshchik:2017kfs}.
At order ${\cal O}(\ep^{-1})$ one finds an equation for $S_2(\gamma,\p)$, which incorporates the first quantum gravitational correction terms \cite{Kiefer:1991}.

In this article we extend the analysis to the case of a scalar-tensor theory. The action \eref{eq.2.1} is rather general, as it involves three arbitrary functions $U(\p)$, $G(\p)$ and $V(\p)$, and covers almost all single field inflationary models in cosmology for different classes of $U$, $G$ and $V$.
Compared to the minimally coupled scalar field there are several differences. 
First, the non-minimal coupling to gravity $U(\p)$ leads to a derivative coupling between the matter and gravitational degrees of freedom, which result in a non-diagonal Wheeler-DeWitt operator \eref{eq.4.7}. Thus, a clear separation of slow and fast degrees of freedom as for the minimally coupled case is no longer available.
Second, in contrast to the minimally coupled case, no constant mass scale $U_0$ is present a priori. 
This make a straightforward application of the semiclassical expansion scheme difficult, as the Born-Oppenheimer approximation relies on a clear separation of slow and fast variables implemented in the Wheeler-DeWitt operator \eref{eq.4.7} by different powers of $\ep$.

We address the problem by the following strategy. It is well known that the scalar-tensor theory \eref{eq.2.1} admits a classically equivalent parametrization, which resembles the action of a scalar field minimally coupled to gravity. This field parametrization is called the Einstein frame (EF). The transition to the EF is achieved by a particular field redefinition $(g,\p)\to(\tilde{g},\tilde{\p})$, which involves a conformal transformation of the $D$-dimensional metric field $g_{\mu\nu}$ and a non-linear field redefinition of the scalar field $\p$. More details can be found in \ref{App:ConformalTransformation} and \ref{App:JFtoEF}. In view of the ADM decomposition, the conformal transformation of the $D$-dimensional metric $g_{\mu\nu}$, induces a corresponding conformal transformation of the geometrical fields $N$, $N_{a}$ and $\gamma_{ab}$ in the canonical theory.
In terms of the EF variables, the Wheeler-DeWitt operator \eref{eq.4.7} becomes diagonal and a clear weighting procedure is available.
Due to the presence of the natural mass scale $U_0$,  we can implement the distinction between heavy and light degrees of freedom by associating each power of $U_0$ with a power of the expansion parameter $\ep$.
After having performed the weighting procedure in the EF, we transform the Wheeler-DeWitt operator \eref{eq.4.7} back to the original Jordan frame (JF) field variables and perform the semiclassical expansion as outlined before.

\subsection{Transition to the Einstein frame}\label{Sec:Einstein frame}
The change of field parametrization from the JF to the EF at the level of the $D$-dimensional covariant Lagrangian \eref{einstein frame} induces a corresponding transformation of the  ADM variables in the canonical formalism 
\begin{equation}
\label{eq.5.6}
\tilde{N}=\left(\frac{U}{U_0}\right)^{\frac{1}{d-1}}N\,,\qquad
\tilde{\gamma}_{ab}=\left(\frac{U}{U_0}\right)^{\frac{2}{d-1}}\gamma_{ab}\,,\qquad 
\frac{\partial\tilde{\p}}{\partial \p}=\left(\frac{U s }{U_0}\right)^{-\sfrac{1}{2}}\,.
\end{equation}
Due to the manifest $d$-dimensional diffeomorphism invariant formulation in terms of the covariant time derivative \eref{eq.2.5}, no explicit factors of the shift vector appear in the formalism.
In terms of the abstract multicomponent configuration space variables $q^{A}$, the transformations \eref{eq.5.6} between the JF and EF can be described by the Jacobi matrices
\begin{eqnarray}
\left(\frac{\partial q^{A}}{\partial \tilde{q}^{B}}\right)=&
\left[
\begin{array}{ccc}
\left(\frac{U}{U_0}\right)^{-\frac{2}{d-1}}\tensor*{\delta}{^{ab}_{cd}}&{}&-\frac{2}{d-1}\frac{U_{1}}{U}\left(\frac{U s }{U_0}\right)^{\sfrac{1}{2}}\gamma_{cd}\\
0&{}&\left(\frac{U s }{U_0}\right)^{\sfrac{1}{2}}
\end{array}
\right]\,,\label{eq.5.7}\\
\left(\frac{\partial \tilde{q}^{B}}{\partial q^{A}}\right)=&
\left[
\begin{array}{ccc}
\left(\frac{U}{U_0}\right)^{\frac{2}{d-1}}\tensor*{\delta}{^{cd}_{ab}}&{}&\frac{2}{d-1}\frac{U_{1}}{U}\tilde{\gamma}_{ab}\\
0&{}&\left(\frac{U s }{U_0}\right)^{-\sfrac{1}{2}}
\end{array}
\right]\label{eq.5.8}\,.
\end{eqnarray}
Note however, that the transformations \eref{eq.5.6} do not simply correspond to a coordinate transformation on configuration space, as beside the transformation of the configuration space variables $q^{A}\to\tilde{q}^{A}$, we also have to take into account the transformation of the lapse function $N\to\tilde{N}$, which, in contrast to $q^{A}=(\gamma_{ab},\p)$ is not a dynamical configuration space variable.
Nevertheless, the inclusion of the lapse function in the definition of the configuration space metric \eref{eq.3.1}, allows to formally write the transformation under \eref{eq.5.6} in the standard covariant form \eref{eq.5.7}, \eref{eq.5.8} as for an ordinary coordinate transformation on configuration space $q^{A}\to \tilde{q}^{A}$, provided that the lapse function is rescaled according to \eref{eq.5.6}.
From the viewpoint of a true coordinate transformation $q^{A}\to\tilde{q}^{A}$, the lapse function $N$ is just a constant. In contrast, for the transformation  between the JF and EF \eref{eq.5.6}, the rescaling of the lapse function has to be taken into account. In particular, it becomes relevant once derivatives of the configuration space metric $\delta_{A}\csm_{BC}$ are transformed from the JF to the EF. This generates additional terms, which will be necessary for the transformation of the Laplace-Beltrami operator from the JF to the EF parametrization. 
The momenta transform covariantly under \eref{eq.5.6},
\begin{equation}
\label{eq.5.9}
\tilde{p}_A=\frac{\partial q^{B}}{\partial \tilde{q}^{A}}\,p_{B}.
\end{equation}
The components of the EF momenta, expressed in terms of the JF momenta, read
\begin{equation}
 \quad\left(\tilde{p}_A\right)= \left(
\begin{array}{c}
\tilde{p}^{ab}_{\tilde{\gamma}}\\
\tilde{p}_{\tilde{\p}}
\end{array}
\right)
=
\left[
\begin{array}{c}
\left(\frac{U}{U_0}\right)^{-\frac{2}{d-1}}\,p^{ab}_{\gamma}\\
\left(\frac{Us}{U_0}\right)^{1/2}\left(p_{\p}-\frac{2}{d-1}\frac{U_{1}}{U}p_{\gamma}\right)
\end{array}
\right]\,.
\end{equation}
Thus, beside multiplicative scaling factors, the EF scalar field momentum $\tilde{p}_{\tilde{\p}}$ is a combination of the JF scalar field momentum $p_{\p}$ and the trace of the JF metric momentum $p_{\gamma}$. This particular combination is a consequence of the original non-minimal coupling responsible for the derivative mixing between the metric and the scalar field degrees of freedom, which becomes manifest in the combined derivative operator \eref{eq.4.8}. In terms of the EF parametrization $(\tilde{\gamma}_{ab},\,\tilde{\p})$, the field content is diagonal, as can be seen explicitly from the diagonal EF configuration space metric $\tilde{\csm}_{AB}$, which together with its inverse $ \tilde{\csm}^{AB}$, can be obtained from the JF configuration space metric \eref{eq.3.1} by
\begin{eqnarray}
\label{eq.5.10}
\tilde{\csm}_{AB}(\tilde{q})&=\frac{\partial q^{C}}{\partial \tilde{q}^{A}}\,\frac{\partial q^{D}}{\partial \tilde{q}^{B}}\,\csm_{CD}(q)\,,\qquad
\tilde{\csm}^{AB}(\tilde{q})=\frac{\partial \tilde{q}^{A}}{\partial q^{C}}\,\frac{\partial \tilde{q}^{B}}{\partial q^{D}}\,\csm^{CD}(q)\,.
\end{eqnarray}
Note that $\tilde{\csm}_{AB}$ and $\tilde{\csm}^{AB}$ transform like ordinary tensors under \eref{eq.5.6}. The explicit expressions for the configuration space metric and its inverse in coordinates $\tilde{q}$ read
\begin{eqnarray}
\left(\tensor{\tilde{\csm}}{_{AB}}\right)
=&\frac{\sig\tilde{\gamma}^{\sfrac{1}{2}}}{\tilde{N}}
\left(
\begin{array}{cc}
-\frac{U_0}{2}\tilde{G}^{abcd} & 0\\
0 & -1
\end{array}
\right),\\
\left(\tensor{\tilde{\csm}}{^{AB}}\right)
=&\frac{\sig\tilde{N}}{\tilde{\gamma}^{\sfrac{1}{2}}}
\left(
\begin{array}{cc}
-\frac{2}{U_0}\tilde{G}_{abcd} & 0\\
0 & -1
\end{array}
\right)\,.
\end{eqnarray}
According to the transformations \eref{eq.5.9} and \eref{eq.5.10}, the quadratic form in the kinetic part of the Hamiltonian transforms as a scalar under \eref{eq.5.6},
\begin{equation}
\tilde{p}_{A}\,\tilde{\csm}^{AB}\,\tilde{p}_{B}=p_{A}\,\csm^{AB}\,p_{B}\,.
\end{equation}
According to the general formula \eref{CTRicciS} and the transformation rules \eref{eq.5.6}, the spatial Ricci scalar in the  EF variables reads 
\begin{equation}
\espricci=\paren{\frac{U}{U_0}}^{-\frac{2}{d-1}}
\left[\spricci
+2\frac{\Delta U}{U}
+\frac{d}{d-1}\frac{\sd_aU\sd^aU}{U^2}
\right].
\label{Ricci}
\end{equation}
According to \eref{EFPot}, the Einstein frame scalar field potential and the spatial derivatives of the scalar field $\p$ transform under \eref{eq.5.6} as
\begin{equation}
 \tilde{\sd}_a\tilde{\p}=\frac{\partial\tilde{\p}}{\partial\p}\,\sd_a\p=\left(\frac{U s }{U_0}\right)^{\sfrac{-1}{2}}\sd_a\p\,,
\qquad
 \tilde{V}=\left(\frac{U}{U_0}\right)^{-\frac{d+1}{d-1}}V\,.
 \label{EFPotDphi}
\end{equation}
Using \eref{Ricci} and \eref{EFPotDphi}, the potential is seen to transform as a scalar under \eref{eq.5.6},
\begin{equation}
\tilde{\csp}=\tilde{\csp}_{\tilde{\gamma}}+\tilde{\csp}_{\tilde{\p}}=\csp_{\gamma}+\csp_{\p}=\csp\,.
\end{equation}
Here, the potentials in the EF parametrization are given by
\begin{equation}
 \tilde{\csp}_{\tilde{\gamma}}=\tilde{N}\tilde{\gamma}^{\sfrac{1}{2}}U_0\tilde{\ricci}^{(d)},\qquad
 \tilde{\csp}_{\tilde{\p}}=\tilde{N}\tilde{\gamma}^{\sfrac{1}{2}}\left(\frac{1}{2}\tilde{\sd}_a\tilde{\p}\tilde{\sd}^a\tilde{\p}+\tilde{V}\right)\,.
\end{equation}
Finally, provided the wave functional $\Psi$ transforms as a scalar $\tilde{\Psi}(\tilde{q})=\Psi(q)$,
the Laplace-Beltrami operator is seen to also transform as a scalar under \eref{eq.5.6},
\begin{eqnarray}
\label{eq.5.11}
\tilde{\square}\tilde{\Psi}(\tilde{q})=&\tilde{\csm}^{AB}(\tilde{q})\tilde{\nabla}_{A}\tilde{\nabla}_{B}\tilde{\Psi}(\tilde{q})=\csm^{AB}(q)\nabla_{A}\nabla_{B}\Psi(q)=\square\Psi(q)\,.
\end{eqnarray}

\subsection{Weighting in the Einstein frame and transformation back to the Jordan frame}
Applying the transformations \eref{eq.5.6}-\eref{eq.5.10} to
\eref{eq.4.6}, we obtain the Wheeler-DeWitt operator in the EF parametrization
\begin{eqnarray}
\fl
\hat{\tilde{\ham}}_{\perp}=\left(\frac{U}{U_0}\right)^{\frac{1}{d-1}}\left\{
\frac{\sig}{\sqrt{\tilde{\gamma}}U_0}\left[\tilde{\dw}_{abcd}\frac{\delta^2}{\delta\tilde{\gamma}_{ab}\delta\tilde{\gamma}_{cd}}
+\frac{(3d+5)(d-2)}{8(d-1)}\tilde{\gamma}_{ab}\frac{\delta}{\delta\tilde{\gamma}_{ab}}\right]\right.\nonumber\\
\left.+U_0\sqrt{\tilde{\gamma}}\espricci+\frac{1}{2}\frac{\sig}{\sqrt{\tilde{\gamma}}}\frac{\delta^2}{\delta\tilde{\p}^2}+\tilde{P}_{\p}\right\}\,.
\end{eqnarray}  
Note that while $\ham$, defined in \eref{eq.3.6}, transforms as a scalar under \eref{eq.5.6}, the Wheeler-DeWitt operator $\ham_\perp$ transforms as a scalar density due to the inverse power of the lapse function in \eref{eq.4.6}.
We separately define the Hamilton operator for the quantum matter degrees of freedom, associated with the EF scalar field
\begin{equation}
\hat{\tilde{\ham}}_{\tilde{\p}}:=\frac{1}{2}\frac{\sig}{\sqrt{\tilde{\gamma}}}\frac{\delta^2}{\delta\tilde{\p}^2}+\tilde{P}_{\tilde{\p}}\,.\label{MatterHphi}
\end{equation} 
In the EF the field content is diagonal and a clear separation between the gravitational degrees of freedom $\tilde{\gamma}_{ab}$ and the scalar degrees of freedom $\tilde{\p}$ is possible.
In the context of the Born-Oppenheimer approximation scheme, this allows us to identify the gravitational variables $\tilde{\gamma}_{ab}$ as the ``slow'' variables and the scalar field $\tilde{\varphi}$ as the ``fast'' variables. Therefore, in the EF there is a clear weighting scheme by associating with each factor of $U_0$ a factor of the dimensionless expansion parameter $\ep$,
\begin{eqnarray}
\fl
\hat{\tilde{\ham}}_{\perp}^{\ep}=\left(\frac{U}{U_0}\right)^{\frac{1}{d-1}}\left\{
\frac{\sig}{\sqrt{\tilde{\gamma}}U_0\ep}\left[\tilde{\dw}_{abcd}\frac{\delta^2}{\delta\tilde{\gamma}_{ab}\delta\tilde{\gamma}_{cd}}
+\frac{(3d+5)(d-2)}{8(d-1)}\tilde{\gamma}_{ab}\frac{\delta}{\delta\tilde{\gamma}_{ab}}\right]\right.\nonumber\\
\left.+U_0\ep\sqrt{\tilde{\gamma}}\espricci+\frac{1}{2}\frac{\sig}{\sqrt{\tilde{\gamma}}}\frac{\delta^2}{\delta\tilde{\p}^2}+\tilde{P}_{\tilde{\p}}\right\}\,.\label{eq.5.12}
\end{eqnarray} 
Note that the overall scaling factor is irrelevant for the weighting process, as only the relative weighting of terms in the Hamiltonian is important. 
The weighted Wheeler-DeWitt operator \eref{eq.5.12} can be transformed back to the JF and reads
\begin{eqnarray}
\fl
\hat{\ham}_{\perp}^{\ep}=\frac{\sig}{\sqrt{\gamma}U\ep }G_{abcd}\frac{\delta^2}{\delta\gamma_{ab}\delta\gamma_{cd}}+\frac{\sig}{\sqrt{\gamma}U\ep }\frac{(3d+5)(d-2)}{8(d-1)}\gamma_{ab}\frac{\delta}{\delta\gamma_{ab}}+\ep P_{\gamma}\nonumber\\
+\frac{\sig}{2}\frac{ s }{\sqrt{\gamma}}\nd^2+\frac{\sig}{4}\frac{ s }{\sqrt{\gamma}}\left(\frac{s_{1}}{s}-\frac{d}{2}\frac{U_{1}}{U}\right)\nd+P_{\p}\label{eq.5.14}\,.
\end{eqnarray}
In the subsequent analysis, in analogy to \eref{MatterHphi}, it will turn out useful to define the Hamilton operator for the fast scalar degrees of freedom
\begin{equation}
 \hat{\ham}_{\rms}:=\frac{\sig}{2}\frac{ s }{\sqrt{\gamma}}\nd^2+\frac{\sig}{4}\frac{ s }{\sqrt{\gamma}}\left(\frac{s_{1}}{s}-\frac{d}{2}\frac{U_{1}}{U}\right)\nd+P_{\p}\,.\label{eq.5.15}
\end{equation}
Note that in contrast to the minimally coupled case, where the scalar field Hamiltonian is free of any factor ordering ambiguities, the covariant Laplace-Beltrami factor ordering \eref{eq.4.5} induces a dependence on the factor ordering in the scalar Hamiltonian \eref{eq.5.15}, reflected by the terms linear in $\nd$. These extra terms can be related directly to the presence of the non-minimal coupling and vanish for $U=U_0$. 
Only part of the gravitational degrees of freedom (the scale part) in $\gamma_{ab}$ mix with the scalar degrees of freedom $\p$ and according to our weighting scheme only these parts are treated as fully quantum in contrast to the remaining gravitational degrees of freedom, which are treated as semiclassical.

\section{Semi-Classical approximation}\label{Sec:SemiClassicalExpansion}
Inserting \eref{eq.5.14}, together with the semiclassical ansatz for the wave functional \eref{eq.5.4} into the Wheeler-DeWitt equation \eref{eq.4.10}, we collect terms of equal order in $\ep$. Demanding that each term vanish separately, we obtain a family of equations for $S_0$, $S_{1}$, $S_{2}$, etc. By truncating at a fixed order in $\ep$ and solving the resulting equations consecutively, starting with the lowest order $S_0$, we can reconstruct the wave functional $\Psi$ in \eref{eq.5.3} within the given accuracy of the approximation. 
In the following subsections, we separately discuss the equations obtained at each order.

\subsection{Order ${\cal O}(\ep^2)$}
At order ${\cal O}(\ep^2)$, we find an equation for $S_0$,
\begin{equation}
\label{LowestOrder}
0=-\frac{1}{2}\frac{\epsilon  s }{\sqrt{\gamma}}(\nd S_0)^2\,.
\end{equation}
Provided $s\neq0$, this is equivalent to
\begin{equation}
\label{eq.6.1}
\nd S_0=\frac{\delta S_0}{\delta\p}-\frac{2}{d-1}\gamma_{ab}\frac{\delta S_0}{\delta\gamma_{ab}}=0\,.
\end{equation}
This implies that $S_{0}(\gamma,\p)=S_0(\tilde{\gamma})$ is only a function of the particular combination
\begin{equation}
\tilde{\gamma}_{ab}=\left(\frac{U}{U_0}\right)^{\frac{2}{d-1}}\gamma_{ab}\,,
\end{equation}
which is nothing but the metric in the EF parametrization \eref{eq.5.6}.

\subsection{Order ${\cal O}(\ep^{1})$}
At order ${\cal O}(\ep^{1})$, we again find an equation for $S_0$, which, after using \eref{eq.6.1}, takes the form
\begin{equation}
\label{eq.6.2}
-\frac{\sig}{U}\frac{\dw_{abcd}}{\sqrt{\gamma}}\frac{\delta S_0}{\delta\gamma_{ab}}\frac{\delta S_0}{\delta\gamma_{cd}}+P_{\gamma}=0\,.
\end{equation}
The structure of \eref{eq.6.2} suggests to define a semiclassical WKB time $\tau$ via
\begin{equation}
\frac{\delta}{\delta\tau}:=-\frac{2\sig}{U}\frac{\dw_{abcd}}{\sqrt{\gamma}}\frac{\delta S_0}{\delta \gamma_{ab}}\frac{\delta }{\delta\gamma_{cd}}\,.\label{eq.6.3}
\end{equation}
In terms of this semiclassical time, \eref{eq.6.2} manifestly acquires the structure of the Hamilton-Jacobi equation for $S_0$ \cite{Peres1962},
\begin{equation}
\label{eq.6.4}
\frac{1}{2}\frac{\delta S_0}{\delta\tau}+P_{\gamma}=0\,.
\end{equation}
The exact Wheeler-DeWitt equation \eref{eq.5.10} is timeless. Therefore, the concept of time only emerges from the semiclassical expansion at the level of the Hamilton-Jacobi equation \eref{eq.6.4}, which, together with the momentum constraint $\hat{\ham}_a\Psi=0$, can be shown to be equivalent to the Einstein equations \cite{Gerlach:1969ph}.\footnote{The quantum momentum constraint equation \eref{eq.4.9} can be expanded in a similar way and ensures the invariance of the wave functional under spatial $d$-dimensional diffeomorphisms order by order \cite{Moncrief:1972cx}.}
Thus, within the semiclassical approximation scheme, the flow of time is associated with the slowly changing background geometry 
$S_0$, which is adiabatically followed by the quantum states of the matter fields.
If we define
\begin{equation}
 \Delta_{0}(\tilde{\gamma}):=\exp\left(-\rmi \ep S_0\right),
\end{equation}
The wave functional to order $\ep$ is simply given by
\begin{equation}
\Psi=\frac{1}{\Delta_{0}(\tilde{\gamma})}\,.
\end{equation}

\subsection{Order ${\cal O}(\ep^{0})$}
At order ${\cal O}(\ep^{0})$, upon using the equation of the previous orders and the definition of semiclassical time \eref{eq.6.3}, we obtain an equation for $S_1$,
\begin{eqnarray}
\fl0=\frac{1}{2}\frac{\sig s }{\sqrt{\gamma}}\left[
-(\nd S_1)^2
+\rmi\nd^2S_1+\frac{\rmi}{2}\left(\frac{ s_1 }{ s }-\frac{d}{2}\frac{U_1}{U}\right)\nd S_1\right]+P_{\p}\nonumber\\
+\frac{\sig}{U\sqrt{\gamma}}\left[
\frac{\rmi}{8}\frac{(3d+5)(d-2)}{d-1}\gamma_{ab}\frac{\delta S_0}{\delta\gamma_{ab}}
+\rmi\dw_{abcd}\frac{\delta^2S_0}{\delta\gamma_{ab}\delta\gamma_{cd}}
\right]+\frac{\delta S_1}{\delta\tau}\,.
\label{expansion3}
\end{eqnarray}
As suggested by the Born-Oppenheimer ansatz, we split $S_{1}$ into a part $\sigma_{1}(\tilde{\gamma})$ which only depends on the background and a part $\Sigma_{1}(\gamma,\p)$, which cannot be reduced further
\begin{equation}
S_1(\gamma,\p)=:\sigma_{1}(\tilde{\gamma})+\Sigma_{1}(\gamma,\p)\,.\label{eq.6.6}
\end{equation}
Inserting \eref{eq.6.6} into \eref{expansion3} and demanding that $\sigma_1$ satisfies the equation
\begin{equation}
\label{eq.6.7}
\rmi\frac{\delta \sigma_1}{\delta\tau}
=\frac{\sig}{U\sqrt{\gamma}}\left[\frac{1}{8}\frac{(3d+5)(d-2)}{d-1}\gamma_{ab}
\frac{\delta S_0}{\delta\gamma_{ab}}
+\dw_{abcd}\frac{\delta^2S_0}{\delta\gamma_{ab}\delta\gamma_{cd}}\right],
\end{equation}
we find an equation for $\Sigma_1$ alone
\begin{equation}
 \frac{\delta\Sigma_1}{\delta\tau}=-\frac{\sig s }{\sqrt{\gamma}}\left[
-\frac{1}{2}(\nd \Sigma_1)^2
+\frac{\rmi}{2}\nd^2\Sigma_1
+\frac{\rmi}{4}\left(\frac{ s_1 }{ s }-\frac{d}{2}\frac{U_1}{U}\right)\nd \Sigma_1\right]-P_{\p}\,.\nonumber\\
\label{eq.6.8}
\end{equation}
If we further define
\begin{equation}
 \psi_{1}:=\exp\left(\rmi \Sigma_1\right),\label{eq.6.9}
\end{equation}
and insert this into \eref{eq.6.8} we obtain a Schr\"odinger equation for $\psi_1$,
\begin{equation}
\label{eq.6.10}
\rmi\frac{\delta\psi_1}{\delta\tau}=\ham_{\mathrm{s}}\,\psi_1.
\end{equation}
The Hamilton operator $\hat{\ham}_{\rms}$ is defined in \eref{eq.5.15}. Moreover, $\sigma_{1}$ is related to the Van Vleck determinant $\Delta_{1}$, which naturally arises in the WKB approximation
\begin{equation}
 \sigma_1=:\rmi \ln\Delta_1(\tilde{\gamma})\,.\label{eq.sigma1}
\end{equation}
Equation \eref{eq.6.10} is a Schr\"odinger equation for the light scalar degree of freedom where the emergent semiclassical time is controlled by the change of the geometry. 
The wave functional up to this order is given by
\begin{equation}
 \Psi=\frac{1}{\Delta_{0}(\tilde{\gamma})\Delta_{1}(\tilde{\gamma})}\psi_1(\gamma,\p)\,.
\end{equation}
At this level of the semiclassical expansion, a notion of unitarity for the light quantum degrees of freedom can be introduced. In this case, unitary evolution of the light degrees of freedom could be defined as the condition
\begin{eqnarray}
	\frac{\delta}{\delta \tau}\langle\psi_1,\psi_1\rangle_{\phi}=0,\label{semiclassunitarity}
\end{eqnarray}
where, in contrast to \eref{eq.4.1}, the inner product $\langle\cdot,\cdot\rangle_{\phi}$ extends only over the light degrees of freedom. 
Note that such a definition of unitarity can at best be a derived semiclassical one, 
its very definition \eref{semiclassunitarity} relies on the notion of a semiclassical time 
$\delta/\delta \tau$ and the derived concept of a Hilbert space of states $\psi_1$ for the light degrees of freedom. 
It might therefore be natural to expect that this concept of semiclassical unitarity breaks down at higher orders of the expansion scheme when quantum gravitational correction terms are included and become relevant.

\subsection{Second order ${\cal O}(\ep^{-1})$}
At ${\cal O}(\ep^{-1})$, we obtain 
\begin{eqnarray}
\fl 0=-\frac{\sig}{U\sqrt{\gamma}}\left[\dw_{abcd}\left(\frac{\delta S_1}{\delta\gamma_{ab}}\frac{\delta S_1}{\delta\gamma_{cd}}
-\rmi\frac{\delta^2S_1}{\delta\gamma_{ab}\delta\gamma_{cd}}\right)-\frac{\rmi}{8}\frac{(3d+5)(d-2)}{d-1}\gamma_{ab}\frac{\delta S_1}{\delta\gamma_{ab}}\right]\nonumber\\
+\frac{\sig s }{\sqrt{\gamma}}\left[\frac{\rmi}{2}\nd^2S_2-\nd S_1\nd S_2+\frac{\rmi}{4}\left(\frac{ s_1 }{ s }-\frac{d}{2}\frac{U_1}{U}\right)\nd S_2\right]+\frac{\delta S_2}{\delta \tau}\,.
\end{eqnarray}
Proceeding in the same way as for $S_{1}$ in \eref{eq.6.6}, we decompose $S_2$ as
\begin{equation}
S_2(\gamma,\p)=:\sigma_{2}(\tilde{\gamma})+\Sigma_{2}(\gamma,\p).
\end{equation}
In analogy to \eref{eq.sigma1}, we define 
\begin{equation}
 \sigma_{2}(\tilde{\gamma})=:\rmi\ln\Delta_{2}
\end{equation}
and choose $\sigma_2$ to be the solution of the equation
\begin{eqnarray}
\label{eq.sigma2}
\fl\rmi\frac{\delta\sigma_2}{\delta\tau}
=\frac{\sig}{U\sqrt{\gamma}}\left[\dw_{abcd}\left(\frac{1}{\Delta_{1}}\frac{\delta^2\Delta_1}{\delta\gamma_{ab}\delta\gamma_{cd}}-2\frac{1}{\Delta_1^2}\frac{\delta \Delta_{1}}{\delta\gamma_{ab}}\frac{\delta \Delta_{1}}{\delta\gamma_{cd}}\right)\right.\nonumber\\
\left.+\frac{1}{8}\frac{(3d+5)(d-2)}{d-1}\gamma_{ab}\frac{1}{\Delta_1}\frac{\delta \Delta_{1}}{\delta\gamma_{ab}}\right].
\end{eqnarray}
The functional $\sigma_{2}(\tilde{\gamma})$ can be interpreted as the second order WKB factor for the heavy degrees of freedom in the Born-Oppenheimer approximation. What remains is an equation for the functional $\Sigma_2(\gamma,\p)$,
\begin{eqnarray}
\label{eq.Sigma2}
\fl\frac{\delta\Sigma_2}{\delta\tau}
=-\frac{\sig s }{\sqrt{\gamma}}\left[\frac{\rmi}{2}\nd^2\Sigma_{2}-\nd\Sigma_{1}\nd\Sigma_{2}+\frac{\rmi}{4}\left(\frac{ s_1 }{s}-\frac{d}{2}\frac{U_1}{U}\right)\nd\Sigma_{2}\right]\nonumber\\
+\frac{\sig}{U\sqrt{\gamma}}\left[\dw_{abcd}\left(\frac{\delta \Sigma_1}{\delta\gamma_{ab}}\frac{\delta \Sigma_1}{\delta\gamma_{cd}}+2\frac{\delta \Sigma_1}{\delta\gamma_{ab}}\frac{\delta \sigma_1}{\delta\gamma_{cd}}
-\rmi\frac{\delta^2\Sigma_1}{\delta\gamma_{ab}\delta\gamma_{cd}}\right)\right.\nonumber\\
\left.-\frac{\rmi}{8}\frac{(3d+5)(d-2)}{d-1}\gamma_{ab}\frac{\delta \Sigma_1}{\delta\gamma_{ab}}\right]\,.\label{eq.6.11}
\end{eqnarray}
In analogy to \eref{eq.6.9}, we define
\begin{equation}
 \psi_{2}:=\exp\left(\rmi \ep^{-1} \Sigma_{2}\right)\,.
\end{equation}
Up to this order of the expansion, the wave functional then reads
\begin{equation}
 \Psi=\frac{1}{\Delta_{0}\Delta_1\Delta_2}\psi_1\psi_2\,.\label{eq.6.12}
\end{equation}
At the previous order, we have found a Schr\"odinger equation \eref{eq.6.10} for $\psi_1$. Since equation \eref{eq.6.11} for $\Sigma_2$ is not very illuminating, we now derive a Schr\"odinger equation for the wave functional corresponding to the fast degrees of freedom, which includes the first quantum gravitational corrections up to order $\ep^{-1}$,
\begin{equation}
 \psi=\psi_1\psi_2=e^{i\left(\Sigma_1+\ep^{-1}\Sigma_2\right)}\,.\label{eq.6.13}
\end{equation}
In order to rewrite \eref{eq.6.11}, we relate derivatives of $\psi$ with those of $\Sigma_1$ and $\Sigma_2$, according to the relation \eref{eq.6.13},
\begin{eqnarray}
 -\rmi\frac{1}{\psi}\frac{\delta \psi}{\delta\tau}&=\frac{\delta \Sigma_1}{\delta \tau}+\ep^{-1}\frac{\delta\Sigma_2}{\delta\tau}\,,\label{eq.6.14}\\
 \nd \Sigma_2&=-\ep \nd \Sigma_1-\rmi\ep \frac{1}{\psi}\nd\psi\,,\label{eq.6.15}\\
 \nd^2\Sigma_2&=-\rmi\ep\frac{1}{\psi}\nd^2\psi-\ep\nd^2\Sigma_1-\rmi\ep\left(\nd\Sigma_1\right)^2-2\rmi\left(\nd\Sigma_1\right)\left(\nd\Sigma_2\right)\label{eq.6.16}\,.
\end{eqnarray}
In the last equation we have neglected terms of ${\cal O}(\ep^{-2})$ arising from the square
 \begin{equation}
  \left(\frac{1}{\psi}\nd\psi\right)^2=-\left(\nd\Sigma_1\right)^2-2\ep^{-1}\left(\nd\Sigma_1\right)\left(\nd\Sigma_2\right)+{\cal O}\left(\ep^{-2}\right)\,.
 \end{equation}
Inserting \eref{eq.6.8} and \eref{eq.6.11} into \eref{eq.6.14} and using \eref{eq.6.15} and \eref{eq.6.16}, we find
\begin{eqnarray}
\fl
\rmi\frac{\delta\psi}{\delta\tau}=\ham_{\rms}\psi-\frac{\sig}{U\ep\sqrt{\gamma}}\left[\dw_{abcd}\left(\frac{2}{\psi_1\Delta_1}\frac{\delta \psi_1}{\delta\gamma_{ab}}\frac{\delta \Delta_1}{\delta\gamma_{cd}}-\frac{1}{\psi_1}\frac{\delta^2\psi_1}{\delta\gamma_{ab}\delta\gamma_{cd}}\right)\right.\nonumber\\
\left.-\frac{1}{8}\frac{(3d+5)(d-2)}{d-1}\gamma_{ab}\frac{1}{\psi_1}\frac{\delta \psi_1}{\delta\gamma_{ab}}\right]\psi\,.\label{eq.6.17}
\end{eqnarray}
This equation is a Schr\"odinger equation for the functional $\psi$ including correction terms indicated by the overall factor of $\ep^{-1}$. 

Let us pause for a moment and summarize the main results of the successive orders in the semiclassical expansion. 
At order $\mathcal{O}\left(\ep^{1}\right)$ we obtained the Hamiltonian-Jacobi equation \eref{eq.6.4} which provides 
a definition semiclassical time \eref{eq.6.3}. At next order, $\mathcal{O}(\ep^{0})$, we obtained a Schr\"odinger 
equation \eref{eq.6.10} for the wave functional of the light degrees of freedom. 
At order $\mathcal{O}\left(\ep^{-1}\right)$, we encountered the first quantum gravitational correction terms \eref{eq.6.17}. 
Ultimately, we are interested in the derivation of the semiclassical branch \eref{eq.5.4} of the full wave functional,
which includes these quantum gravitational corrections. 
The successive solution of the equations \eref{eq.6.4}, \eref{eq.6.7}, \eref{eq.6.8}, \eref{eq.sigma2} and \eref{eq.Sigma2} 
allows to construct this wave functional \eref{eq.5.4} up to the required order. 
In principle, we could therefore finish our analysis at that point. 

However, in the following section we reformulate \eref{eq.6.17} to obtain a clearer interpretation of the structure of the 
quantum gravitational correction terms. 
Moreover, since the main motivation of this work is to study the impact of the non-minimal coupling on the quantum 
gravitational correction terms, we would like to express \eref{eq.6.17} in a from in which we can compare it to the minimally 
coupled case analyzed in \cite{Kiefer:1991}.

\subsection{Representation of the correction terms}
Provided the equations for $S_0$ and $S_1$ have been solved, solving equation \eref{eq.6.10} and \eref{eq.6.11} is sufficient to determine the wave functional up to order ${\cal O}\left(\ep^{-1}\right)$ and therefore the final result \eref{eq.6.12}. The form of the correction terms in \eref{eq.6.17} is however not very illuminating. In order to write these terms in a more transparent form and to compare them with the result for the minimally coupled scalar field, we repeat here the same steps as in \cite{Kiefer:1991}.

The correction terms in \eref{eq.6.17} can be decomposed into contributions orthogonal and tangential to the surfaces of constant $S_0$.
The first step is to decompose $\delta\psi_1/\delta\gamma_{ab}$ into its components tangential and orthogonal to the surfaces of constant $S_0$. Note that $\delta S_0/\delta \gamma_{ab}$ corresponds to the $\gamma_{ab}$-component of the configuration space gradient of $S_0$ and therefore to the $\gamma_{ab}$-components of a natural covector $\delta S_0/\delta q^{A}$. With respect to the vector product on configuration space, we calculate the norm
\begin{eqnarray}
 \left\langle\frac{\delta S_0}{\delta q},\frac{\delta S_{0}}{\delta q}\right\rangle&:=\frac{\delta S_0}{\delta q^{A}}\csm^{AB}\frac{\delta S_0}{\delta q^{B}}=\frac{\delta S_0}{\delta\gamma_{ab}}\left(-\frac{2\sig N}{U}\dw_{abcd}\right)\frac{\delta S_0}{\delta\gamma_{cd}}\nonumber\\
 &=-2\csp_{\gamma}=\sig\langle n,n\rangle\,,
\end{eqnarray}
where we have used \eref{eq.6.1}, \eref{eq.6.3} and the Hamilton-Jacobi equation \eref{eq.6.4} in the last step.
The unit normal covector $n_{A}$ is therefore defined as
\begin{equation}
 n_{A}:=\sig\frac{\delta S_0}{\delta q^{A}}\left\langle \frac{\delta S_0}{\delta q^{A}},\frac{\delta S_0}{\delta q^{A}}\right\rangle^{-\sfrac{1}{2}} =\sig(-2\sig\csp_{\gamma})^{-\sfrac{1}{2}} \frac{\delta S_0}{\delta q^{A}}\,,
\end{equation}
and the normal vector $n^{A}$ by
\begin{equation}
 n^{A}=\sig(-2\sig\csp_{\gamma})^{-\sfrac{1}{2}}\csm^{AB}\frac{\delta S_0}{\delta q^{B}}\,.\label{eq.6.18}
\end{equation}
Note that for $U=U_0$ and $R^{(d)}=0$ the construction of the normal vector is not possible and one would have to resort to the corrections in the form as they appear in equation \eref{eq.6.17}.
We introduce projection operators normal and tangential to the surface of constant $S_0$,
\begin{equation}
\tensor{\Pi}{_{\perp}_{A}^{B}}=\sig n_{A}n^{B},\qquad \tensor{\Pi}{_{\parallel}_{A}^{B}}=\delta^{B}_{A}-\sig n_{A}n^{B}\,.
\end{equation}
The normal projection of $\delta\psi_1/\delta q^{A}$ then reads explicitly
\begin{eqnarray}
 \tensor{\Pi}{_{\perp}_A^B}\,\frac{\delta \psi_1}{\delta q^B}&=\sig n_An^B\frac{\delta \psi_1}{\delta q^B}=\left(-2\csp_{\gamma}\right)^{-1}\left\langle\frac{\delta S_0}{\delta q},\frac{\delta \psi_1}{\delta q}\right\rangle\frac{\delta S_0}{\delta q^{A}}\nonumber\\
 &=-\frac{1}{2P_{\gamma}}\frac{\delta \psi_1}{\delta \tau}\frac{\delta S_0}{\delta q^A}=\frac{\rmi}{2P_{\gamma}}\left(\hat{\ham}_{\rms}\psi_{1}\right)\frac{\delta S_0}{\delta q^{A}}\,,
\end{eqnarray}
where we have used 
\begin{equation}
 \left\langle\frac{\delta S_0}{\delta q},\frac{\delta \psi_1}{\delta q}\right\rangle=-\frac{2\sig N}{U}\frac{\dw_{abcd}}{\sqrt{\gamma}}\frac{\delta\psi_1}{\delta\gamma_{ab}}\frac{\delta S_0}{\delta\gamma_{cd}}\,.\label{eq.6.19}
\end{equation}
The $\gamma_{ab}$-component of the normal projection of $\delta\psi_{1}/\delta q^A$ is then given by
\begin{eqnarray}
\left(\Pi_{\perp}\frac{\delta\psi}{\delta\gamma}\right)^{ab}=&\frac{\rmi}{2P_{\gamma}}\left(\hat{\ham}_{\rms}\psi_{1}\right)\frac{\delta S_0}{\delta\gamma_{ab}}\,,
\end{eqnarray}
where we have used \eref{eq.6.18} and \eref{eq.6.19} in the last step.
In the same way, we introduce the $\gamma_{ab}$-component of the tangential projection of $\delta\psi_{1}/\delta q^A$,
\begin{equation}
 T^{ab}:=\left(\Pi_{\parallel}\frac{\delta\psi}{\delta\gamma}\right)^{ab}\,,\qquad \left\langle T,n\right\rangle
 =0\,.
\end{equation}
This allows to decompose $\delta \psi_{1}/\delta\gamma_{ab}$ into its normal and tangential parts
\begin{equation}
\label{eq.6.21}
 \frac{\delta \psi_1}{\delta\gamma_{ab}}=\frac{\rmi}{2P_{\gamma}}\left(\hat{\ham}_{\rms}\psi_{1}\right)\frac{\delta S_0}{\delta\gamma_{ab}}+T^{ab}\,.
\end{equation}
Note that we do not specify $T_{ab}$ explicitly as we will not need the tangential contributions in the correction terms, see \cite{Kiefer:1991}. 
Differentiating \eref{eq.6.21} with respect to \(\gamma_{cd}\), we obtain
\begin{eqnarray}
\label{eq.6.22}
\fl
\frac{\delta^2\psi_1}{\delta\gamma_{ab}\delta\gamma_{cd}}
=\frac{\rmi}{2P_\gamma}\left[\frac{\delta S_0}{\delta\gamma_{cd}}\left(-\frac{1}{P_\gamma}\frac{\delta P_\gamma}{\delta\gamma_{ab}}\hat{\ham}_\rms
+\frac{\delta\hat{\ham}_\rms}{\delta\gamma_{ab}}
+\hat{\ham}_\rms\frac{\delta}{\delta\gamma_{ab}}\right)+\frac{\delta^2S_0}{\delta\gamma_{ab}\delta\gamma_{cd}}\hat{\ham}_\rms\right]\psi_1\nonumber\\
+\frac{\delta T^{cd}}{\delta\gamma_{ab}}\,.
\end{eqnarray}
Inserting \eref{eq.6.21} and \eref{eq.6.22} into \eref{eq.6.17} and making use of \eref{eq.6.7}, we split the correction terms into two parts
\begin{eqnarray}
\fl
B_n+B_t=-\frac{\sig}{U\lambda\sqrt{\gamma}\psi_1}\left[\dw_{abcd}\left(\frac{2}{\Delta_1}\frac{\delta\psi_1}{\delta\gamma_{ab}}\frac{\delta\Delta_1}{\delta\gamma_{cd}}-\frac{\delta^2\psi_1}{\delta\gamma_{ab}\delta\gamma_{cd}}\right)\right.\nonumber\\
\left.-\frac{1}{8}\frac{(3d+5)(d-2)}{d-1}\gamma_{ab}\frac{\delta\psi_1}{\delta\gamma_{ab}}\right]\,.
\end{eqnarray}
The tangential part of the corrections reads
\begin{eqnarray}
\fl
B_t=
-\frac{\sig}{U\ep\sqrt{\gamma}\psi_1}
&\left[\dw_{abcd}
\left(\frac{2}{\Delta}\frac{\delta\Delta_1}{\delta\gamma_{ab}}
     -\frac{\delta }{\delta\gamma_{ab}}
     -\frac{\rmi}{2P_\gamma}\frac{\delta S_0}{\delta\gamma_{ab}}\hat{\ham}_\rms \right)T^{cd}\right.\nonumber\\
&
\left.-\frac{1}{8}\frac{(3d+5)(d-2)}{d-1}\gamma_{ab}T^{ab}\right]\,.
\end{eqnarray}
The orthogonal part of the corrections reads
\begin{equation}
B_n=
-\frac{1}{4\ep\psi_1}\left[\frac{1}{P_\gamma}\frac{\sig}{U}\frac{\dw_{abcd}}{\sqrt{\gamma}}\frac{\delta S_0}{\delta\gamma_{ab}}
                        \hat{\ham}_\rms\left(\frac{\delta S_0}{\delta\gamma_{cd}}\frac{1}{P_\gamma}\hat{\ham}_{\rms}\right)
                       +\rmi\frac{\delta}{\delta\tau}\left(\frac{1}{P_\gamma}\hat{\ham}_\rms\right)\right]\psi_1.
\end{equation}
Notice, that due to \eref{eq.6.13} the derivatives of $\psi_1$ can be replaced by derivatives of $\psi$ at this order of the expansion
\begin{equation}
\frac{\nd \psi}{\psi}=\frac{\nd \psi_1}{\psi_1}+{\cal O}\left(\frac{1}{\ep}\right).
\end{equation}
The correction terms can therefore be expressed in the form of a corrected Schr\"odinger equation for $\psi$.
In contrast to the normal contributions, which are proportional to $\delta S_0/\delta \gamma_{ab}$ and therefore determined by the previous orders of the expansion, the tangential contributions are undetermined. Neglecting the tangential terms as in \cite{Kiefer:1991}, we obtain the quantum gravitationally corrected
Schr\"odinger equation
\begin{equation}
\rmi\frac{\delta\psi}{\delta\tau}
=\hat{\ham}_\rms\psi
-\frac{1}{4\lambda}\left[\frac{1}{P_{\gamma}}\frac{\epsilon}{U}\frac{\dw_{abcd}}{\sqrt{\gamma}}
\frac{\delta S_0}{\delta\gamma_{ab}}
\hat{\ham}_{\rms}   
\frac{\delta S_0}{\delta\gamma_{cd}}
\frac{1}{P_\gamma}
\hat{\ham}_\rms
+\rmi\frac{\delta}{\delta\tau}
\left(\frac{1}{P_\gamma}\hat{\ham}_{\rms}\right)\right]\psi\,.\label{eq.6.23}
\end{equation}
Compared to the analysis for a minimally coupled scalar field, performed in \cite{Kiefer:1991}, there are several differences, which we discuss in the remaining part of this section.

The last term in \eref{eq.6.23} has the same structure as in the minimally coupled case and has been associated with a unitarity violating term in \cite{Kiefer:1991}.

The appearance of the unitarity violating term in \eref{eq.6.23} can be traced back to the use of the uncorrected Schr\"odinger equation \eref{eq.6.10} in the process of reformulating the result \eref{eq.6.17} into a form that resembles a Schr\"odinger equation for $\psi$. Moreover, the apparent unitarity violating is to be understood here at the semiclassical level in the sense of \eref{semiclassunitarity}. Since this semiclassical concept of unitarity can be at most an effective one, which emerges from the semiclassical expansion itself, it might be expected to break down once quantum gravitational corrections become relevant. A way to deal with the unitarity violating term in \eref{eq.6.23} would be to formally absorb it by a redefinition of the semiclassical time $\tau$ (see e.g. \cite{Kiefer:2018xfw}). It can be seen that this would correspond to the inclusion of backreaction terms of the light degrees of freedom to the slow degrees of freedom. More precisely these backreaction terms would modify the background $S_{0}$ and therefore the Hamilton-Jacobi equation \eref{eq.6.4} which defines the semiclassical time \eref{eq.6.3}. In the context of a reduced minisuperspace model of a minimally coupled scalar field, the authors in \cite{Bertoni:1996ms} find that the inclusion of backreaction terms leads to a unitary semiclassical evolution. In this work we neglect backreaction effects as we are mainly interested in analyzing the impact of the non-minimal coupling on the structure of the first quantum gravitational correction terms. In particular, we would like to compare our results to the semiclassical expansion of the Wheeler-DeWitt for a minimally coupled scalar field, which was treated in \cite{Kiefer:1991} where backreaction terms where neglected.

In the following, we therefore focus on the first term in \eref{eq.6.23}, which entails the relevant quantum gravitational corrections.
This term is not of the same form as in the minimally coupled case in \cite{Kiefer:1991}.
The reason for this is that the Hamiltonian $\hat{\ham}_{\rms}$ does not commute with $\delta S_0/\delta \gamma_{cd}$ and $P_{\gamma}$. If we nevertheless try to arrange the correction terms in a form similar to those in \cite{Kiefer:1991}, we have to take into account additional commutators, which we discuss below.
In particular, commuting $\delta S_0/\delta\gamma_{ab}$ through $\hat{\ham}_{\rms}$ allows to make use of the Hamilton-Jacobi equation \eref{eq.6.4} and thereby to eliminate all occurrences of $\delta S_0/\delta\gamma_{ab}$. 
Commuting $\delta S_0/\delta\gamma_{ab}$ through $\hat{\ham}_{s}$, we find 
\begin{eqnarray}
\fl
\hat{\ham}_\rms\left(\frac{\delta S_0}{\delta\gamma_{ab}}\frac{1}{P_\gamma}\hat{\ham}_\rms\psi\right)
=\frac{\delta S_0}{\delta\gamma_{ab}}\left\{\hat{\ham}_\rms\frac{1}{P_\gamma}\hat{\ham}_\rms\psi
+\frac{2}{d-1}\frac{\sig s }{\sqrt{\gamma}}\frac{U_1}{U}\nd\frac{1}{P_\gamma}\hat{\ham}_\rms\psi\right.\nonumber\\
\left.
+\frac{1}{d-1}\frac{\sig s }{\sqrt{\gamma}}
\left[\frac{U_2}{U}+\frac{1}{2}\frac{ s_1 U_1}{ s  U}-\frac{d^2+3d-12}{4(d-1)}\left(\frac{U_1}{U}\right)^2\right]
\frac{1}{P_\gamma}\hat{\ham}_\rms\psi\right\}\,.
\end{eqnarray}
Inserting this into \eref{eq.6.23}, the corrected Schr\"odinger equation for $\psi$ acquires the form
\begin{eqnarray}
\fl
\rmi\frac{\delta\psi}{\delta\tau}=\left(1-\frac{f(\p)}{\lambda P_\gamma}\right)\hat{\ham}_\rms\psi\nonumber\\
-\frac{1}{4\lambda}\left[\hat{\ham}_\rms\frac{1}{P_\gamma}\hat{\ham}_\rms
+\frac{2\left[\delta^{(d)}(0)\right]}{d-1}\frac{\sig s }{\sqrt{\gamma}}\frac{U_1}{U}\nd\frac{1}{P_\gamma}\hat{\ham}_{\rms} 
+\rmi\frac{\delta}{\delta\tau}\left(\frac{1}{P_\gamma}\hat{\ham}_\rms\right)\right]\psi\,.
\label{6.24}
\end{eqnarray}
Notice that we have collected the correction terms from last line in \eref{6.24} in the function $f(\p)$ and restored explicit factors of $\delta^{(d)}(0)$,
\begin{eqnarray}
f(\p)=\frac{\left[\delta^{(d)}(0)\right]^2}{16(d-1)}\frac{\sig s }{\sqrt{\gamma}}
\left[4\frac{U_2}{U}+2\frac{ s_1 U_1}{ s  U}-\frac{d^2+3d-12}{d-1}\left(\frac{U_1}{U}\right)^2\right]\,.
\end{eqnarray}
Thus, in contrast to the minimally coupled case, singular factors of $\delta^{d}(0)$ enter the final result from two different sources -- due to the Laplace-Beltrami factor ordering \eref{eq.4.5} and due to the non-commutativity of derivatives in $\hat{\ham}_{\rms}$ with background quantities. 

Adopting a regularization as mentioned in \cite{DeWitt:1967yk}, where the $\delta^{(d)}(0)$ contributions are regularized to zero, effectively corresponds to omitting all derivative terms that do not act on the wave functional. In this case, the matter Hamiltonian $\hat{\ham}_{\rms}$ can be commuted to the very right
\begin{eqnarray}
\rmi\frac{\delta\psi}{\delta\tau}=&\hat{\ham}_{\rms}^{\mathrm{P}}\psi-\frac{1}{4}\left[\frac{1}{P_{\gamma}}\left(\hat{\ham}_{\rms}^{\mathrm{P}}\right)^2 
+\rmi\frac{\delta}{\delta\tau}\left(\frac{1}{P_\gamma}\hat{\ham}_{\rms}^{\mathrm{P}}\right)\right]\psi\,,\label{eq.6.25}
\end{eqnarray}
and the kinetic term of $\hat{\ham}_{\rms}$ reduces to its principal part
\begin{eqnarray}
\hat{\ham}_{\rms}^{\mathrm{P}}:=\frac{\sig}{2}\frac{ s }{\sqrt{\gamma}}\nd^2+P_{\p}\,.
\label{eq.6.26}
\end{eqnarray}
It is understood that the $\p$ and $\gamma_{ab}$ derivatives in $\nd$ only act on the wave functional $\psi$ in \eref{eq.6.25}.
The form of the correction terms in \eref{6.24} then features the same structure as in the minimally coupled case -- the only difference is that the scalar Hamilton operator \eref{eq.6.26} replaces the matter Hamiltonian of the minimally coupled scalar field 
\begin{equation}
\hat{\ham}_{\p}=\frac{\sig}{2\sqrt{\gamma}}\frac{\delta^2}{\delta\p^2}+P_{\p}\,.\label{eq.6.27}
\end{equation}
The difference between \eref{eq.6.26} and \eref{eq.6.27} consists of the generalized derivative operator $\nd$ instead of the simple $\delta/\delta\p$ and the overall factor $ s $ in the kinetic part of \eref{eq.6.26}.

We discuss the effects of these differences in the context of the cosmological model of Higgs inflation in $d=3$,
where $\p$ is associated with the Standard Model Higgs boson and for which the arbitrary functions in the general scalar-tensor theory \eref{eq.2.1} acquire the particular form
\begin{eqnarray}
U(\p)=\frac{1}{2}\left(M_{\mathrm{P}}^2+\xi\p^2\right),\qquad G(\p)=1,\qquad V(\p)=\frac{\lambda}{4}\left(\p^2-\nu^2\right)^2\,.\label{eq.6.28}
\end{eqnarray}
Here $\xi$ is the non-minimal coupling constant, $\lambda$ the quartic Higgs self-interaction and $\nu\simeq246$ GeV the electroweak symmetry breaking scale.
The form of the non-minimal coupling function $U(\p)$ in \eref{eq.6.28}, clearly shows that the relevant parameter is the dimensionless combination
\begin{equation}
x:=\frac{\sqrt{\xi}\p}{M_{\rm P}}\,.
\end{equation}
Let us investigate two regions in configuration space, corresponding to the asymptotic regimes $x\ll1$ and $x\gg1$. For a weak non-minimal coupling and $x\ll1$ the first term of $U$ in \eref{eq.6.28} dominates and we expect to recover the minimally coupled case.
Indeed, for the functions \eref{eq.6.28}, the function $ s $, defined in \eref{eq.3.3}, can be expressed in terms of $x$ and $\xi$ as
\begin{equation}
 s =\frac{1+x^2}{1+(1+6\xi)x^2}\,.\label{eq.6.29}
\end{equation}
Clearly, for $x\ll1$, the function $ s $ tends to one as \eref{eq.6.29} reduces to
\begin{equation}
 s =1+{\cal O}\left(x^2\right)\,.\label{eq.6.30}
\end{equation}
Next, let us analyze the derivative $\nd$ in this limit
\begin{equation}
\nd=\frac{\delta}{\delta\p}-\frac{x}{1+x^2}\sqrt{\frac{4\xi}{M_{\mathrm{P}}}}\gamma_{ab}\frac{\delta}{\delta\gamma_{ab}}\,.
\end{equation}
For $x\ll1$, the $\gamma_{ab}$ derivative is suppressed by $x\sqrt{\xi}/M_{\mathrm{P}}$,
\begin{equation}
\nd=\frac{\delta}{\delta\p}+{\cal O}\left(x\right)\,.\label{eq.6.31}
\end{equation}
Thus, in view of \eref{eq.6.30} and \eref{eq.6.31}, in the limit $x\ll1$, the scalar matter Hamilton operator \eref{eq.6.26} reduces to the matter Hamilton operator \eref{eq.6.27} for the minimally coupled scalar field. This relation is in fact a required consistency condition, as for $\xi\to0$ we must recover the minimally coupled case.  

Next, let us investigate the $x\gg1$ case. In this case the function $ s $ reduces to
\begin{equation}
 s =\frac{1}{1+6\xi}+{\cal O}\left(\frac{1}{x^2}\right)\approx\frac{1}{6\xi}\ll1,
\end{equation}
where we have assumed a large non-minimal coupling $\xi\gg1$. Thus, for a strong non-minimal coupling $\xi$, the function $s$ leads to a strong overall suppression of the kinetic terms in \eref{eq.6.26} in the $x\gg1$ regime.
Nevertheless, the $\gamma_{ab}$ derivative in $\nd$ is still suppressed by a factor of $\sqrt{\xi}/M_{\mathrm{P}}x$ for $x\gg1$,
\begin{equation}
\nd=\frac{\delta}{\delta\p}+{\cal O}\left(\frac{1}{x}\right)\,.\label{ndlargex}
\end{equation}
Therefore, independent of $\xi$, the metric derivatives in $\nd$ are suppressed in both cases for $x\ll 1$ and $x\gg 1$. This behavior can be traced back to the function $x/(1+x^2)$, which tends to zero for $x\ll 1$ and $x\gg1$ and has a global maximum (for $x\geq0$) in between the two asymptotic regimes at $x=1$. Thus the effect of the $\gamma_{ab}$ derivatives is strongest for $\varphi=M_{\mathrm{P}}/\sqrt{\xi}$ corresponding to $x=1$.
The suppression of the kinetic term in \eref{eq.6.26} by the function $ s $ for a strong non-minimal coupling might be interpreted as the analogue of the suppression mechanism of the Higgs propagator found in the perturbative covariant approach to Higgs inflation \cite{Bezrukov:2007ep, Barvinsky:2008ia,DeSimone:2008ei,Bezrukov:2008ej,Barvinsky:2009ii,Barvinsky:2009fy,Bezrukov:2010jz}.

\section{Conclusions}\label{Sec:Conclusions}
We have performed the canonical quantization of a general scalar-tensor theory. We have derived the Wheeler-DeWitt equation and performed its semiclassical expansion. At the lowest orders of this expansion we have recovered the classical theory. At the higher orders of the expansion, we have found that the semiclassical wave functional satisfies a Schr\"odinger equation, which includes the first quantum gravitational correction terms.
Throughout the paper, we have formally treated the configuration space as a differentiable manifold and derived all the associated geometrical tensors, including the scalar Ricci curvature of configuration space. In particular, we have found that, in contrast to pure gravity, in the case of a non-minimally coupled scalar field, the signature of the configuration space metric depends on the signature of spacetime. This might have interesting consequences regarding the hyperbolicity properties of the Wheeler-DeWitt operator \cite{Kiefer:1989km}. 
As required for consistency, at each step of the calculation we recover the results for a minimally coupled scalar field with a canonically normalized kinetic term obtained in \cite{Kiefer:1991} by setting $U(\p)=U_0$ and $G=1$. In contrast, for arbitrary field dependent functions $U(\p)$ and $G(\p)$, the canonical quantization and the subsequent semiclassical expansion lead to essential differences compared to the minimally coupled case -- both technical and conceptual.

In particular, the non-minimal coupling $U(\p)$ leads to a mixing between the gravitational and scalar field momenta. This intertwining of gravitational and scalar field degrees of freedom makes it difficult to separate heavy from light degrees of freedom in the multicomponent configuration space. While this might not pose a problem in principle, at the level of the exact Wheeler-DeWitt equation, it complicates the semiclassical expansion.
The semiclassical expansion is based on the Born-Oppenheimer approximation, which in turn requires a clear separation of heavy and light degrees of freedom. Moreover, in contrast to the minimally coupled case, where the Planck mass $U_0=M_{\mathrm{P}}^2/2$ serves as a natural indicator for the heavy degrees of freedom, in general no such constant scale is present for an arbitrary non-minimal coupling function $U(\p)$.

Practically, the semiclassical expansion of the Wheeler-DeWitt equation requires a relative weighting between individual terms in the Wheeler-DeWitt operator by different powers of $\ep$ which implements the distinction between heavy and light degrees of freedom. A concrete weighting procedure in case of a non-minimal coupling is therefore difficult, as the Wheeler-DeWitt operator is non-diagonal. In order to nevertheless obtain a consistent and feasible weighting scheme, we have performed a transformation to the Einstein frame, in which the Wheeler-DeWitt operator is diagonal. In the Einstein frame, the distinction between gravitational and scalar field degrees of freedom is transparent and a clear weighting can be performed by associating the Einstein frame metric field with the heavy degrees of freedom and the scalar field with the light degrees of freedom. Once the weighting has been implemented, the weighted Wheeler-DeWitt operator can then be transformed back to the original Jordan frame variables, and the semiclassical expansion can be carried out.   

The justification of this procedure relies on the covariant Laplace-Beltrami ordering. On the basis of covariant perturbative one-loop calculations \cite{Barvinsky:1993zg, Shapiro:1995yc, Steinwachs:2011zs, Kamenshchik:2014waa, wefR}, the quantum equivalence between different parametrizations of scalar-tensor theories \cite{Kamenshchik:2014waa} and the equivalence between $f(R)$-gravity and its reformulation as a scalar-tensor theory has been investigated for the one-loop divergences on a general background manifold in \cite{weframefr}; see also a related discussion about the equivalence of the effective action in the context of Einstein spaces in \cite{Ohta:2017trn}. There, it has been found that the classical equivalence is broken by off-shell contributions but is restored once the equations of motions have been used. In the geometrical treatment of the configuration space, the quantum equivalence between the Jordan frame and Einstein frame in the non-perturbative canonical theory can be realized, at least formally, by the covariant Laplace-Beltrami factor ordering in the Wheeler-DeWitt operator. It would be interesting to investigate whether this quantum equivalence also holds 
between $f(R)$-gravity and its scalar-tensor formulation in quantum geometrodynamics.

For the minimally coupled scalar field case, the final result for the corrected Schr\"odinger equation is independent of the factor ordering in the kinetic part of the Wheeler-DeWitt operator. In contrast, for the general scalar-tensor theory \eref{eq.2.1}, the factor ordering is determined by the Laplace-Beltrami operator and ultimately enters the corrected Schr\"odinger equation. The additional terms, which arise in the Laplace-Beltrami factor ordering, correspond to lower order derivative terms in the Wheeler-DeWitt operator and involve delta functions evaluated at the same point. In addition, compared to the minimally coupled case, extra commutator terms have to be taken into account in the corrected Schr\"odinger equation, which also carry factors of $\delta^{(d)}(0)$.
These singular delta functions need to be regulated.
Adopting a regularization scheme, in which operators at the same point commute \cite{DeWitt:1967yk}, the corrected Schr\"odinger equation acquires the same form as for the minimally coupled scalar field, but with the minimally coupled scalar Hamilton operator \eref{eq.6.27} replaced by the non-minimal scalar Hamilton operator \eref{eq.6.26}. The kinetic term of the latter involves derivatives with respect to the scalar field as well as derivatives with respect to gravitational metric. Moreover,  the structure of the quantum gravitational correction terms in the case of non-minimal coupling shows additional interesting differences compared to the minimally coupled case. We have investigated the nature of these differences for the specific model of non-minimal Higgs inflation \cite{Bezrukov:2007ep, Barvinsky:2008ia,DeSimone:2008ei,Bezrukov:2008ej,Barvinsky:2009ii,Barvinsky:2009fy,Bezrukov:2010jz}.
In particular, in the regime of a strong non-minimal coupling, the kinetic part of the scalar Hamilton operator \eref{eq.5.15} was shown to be strongly suppressed. A similar effect has been found in the model of Higgs inflation, where, in the presence of a strong non-minimal coupling, the Higgs propagator is suppressed at high energies \cite{DeSimone:2008ei,Bezrukov:2008ej,Barvinsky:2009ii}.

It would be interesting to explore the features of the quantum gravitational correction terms and the influence of the non-minimal coupling in the canonical theory by applying it to a homogeneous and isotropic cosmological background -- including perturbations. 
In contrast to the weighting scheme adopted in the present work, in the semiclassical expansion of such a cosmological minisuperspace model, the homogeneous scalar field $\p(t)$ and the scale factor $a(t)$ could be identified as the slow variables and both be treated on equal footing, while the fast degrees of freedom are provided in a natural way by the inhomogeneous cosmological perturbations.
Most importantly, in a similar way as for a minimally coupled scalar field \cite{Kiefer:2011cc, Bini:2013fea, Kamenshchik:2013msa, Kamenshchik:2014kpa, Brizuela:2015tzl, Brizuela:2016gnz}, such a cosmological application would allow the estimation of the effect of a non-minimal coupling on quantum gravitational contributions to the power spectrum of the cosmic microwave background radiation.

\section*{Acknowledgements}
M.L.W. was supported by the Research Training Group GRK 2044 of the German Research Foundation (DFG).

\appendix

\section{Geometry of configuration space}\label{App:GeometryConfigurationSpace}
In this appendix the configuration space is formally considered as an (infinite dimensional) differentiable manifold with line element
\begin{equation}
\rmd s^2=\int\rmd^dx\,\csm_{AB}\,\rmd q^{A}\rmd q^{B}.
\end{equation}
For all expressions in the following subsections we suppress the factors of $\delta^{(d)}(0)$ arising from functional differentiation at the same point.

\subsection{Christoffel symbols configuration space}
The Christoffel symbol constructed from the configuration space metric $\csm_{AB}$ reads
\begin{equation}
\tensor{\cschrist}{^C_{AB}}=\frac{1}{2}\csm^{CD}\paren{\delta_A\csm_{DB}+\delta_B\csm_{AD}-\delta_D\csm_{AB}}\,.
\end{equation}
For the explicit components of the Christoffel symbol, we find
\begin{eqnarray}
\tensor{\cschrist}{^\p_{\p\p}}
&=-\frac{1}{2}\paren{\frac{ s_1 }{ s }
+\frac{2d}{(d-1)^2}\frac{ s  U_{1}^3}{U^2}-\frac{d}{d-1}\frac{U_{1}}{U}},\label{ccsm1}\\
\tensor{\cschrist}{^{\p ab}_\p}
&=\frac{1}{4}\paren{1-\frac{2}{d-1}\frac{ s  U_{1}^2}{U}}\gamma^{ab},\\
\tensor{\cschrist}{^{\p abcd}}
&=\frac{1}{4}\frac{ s  U_{1}}{d-1}G^{abcd},\\
\tensor{\cschrist}{_{ab\p\p}}
&=\frac{1}{2(d-1)}\left[\frac{1}{ s  U}-\frac{4d}{d-1}\left(\frac{ U_{1}}{U}\right)^2\right.\nonumber\\
&\qquad\left.  
+2\frac{ s_1 }{ s }\frac{U_{1}}{U}+4\frac{U_{2}}{U}
+\frac{4d}{(d-1)^2}\frac{ s   U_{1}^4}{U^3}\right]\gamma_{ab},\\
\tensor{\cschrist}{_{ab\p}^{cd}}  &=\frac{1}{2}\frac{U_{1}}{U}
\left[\delta_{ab}^{cd}-\frac{1}{d-1}\left(1-\frac{2}{d-1}\frac{ s  U_{1}^2}{U}\right)
\gamma_{ab}\gamma^{cd}\right],\\
\tensor{\cschrist}{_{ab}^{ef}^{cd}} &=\frac{1}{2}\delta_{ab}^{((cd}\gamma_{\phantom{\dagger}}^{ef))}
-\delta_{ab}^{(c(e}\gamma_{\phantom{\dagger}}^{f)d)}\nonumber\\
&+\frac{1}{4(d-1)}\paren{1-\frac{2}{d-1}\frac{ s  U_{1}^2}{U}}
\gamma_{ab}\dw^{cdef}\label{ccsm6}.
\end{eqnarray}
By construction, the Christoffel symbol $\tensor{\Gamma}{^{K}_{AB}}$ satisfies the metric compatibility condition
\begin{equation}
\nabla_A\csm_{BC}=0.
\end{equation}
We confirmed this explicitly as a consistency check for the components of the Christoffel symbol \eref{ccsm1}-\eref{ccsm6}.

\subsection{Metric determinant}
Since the DeWitt metric $\gamma^{\sfrac{1}{2}}G^{abcd}$ is invertible for non-singular metrics, 
the determinant of the configuration space metric \eref{eq.3.1} can be calculated by
\begin{equation}
\label{detcsm0}
\csm:=\det\left(\csm_{AB}\right)=
\det\left(\csm_{\gamma\gamma}\right)\det\paren{\csm_{\p\p}-\csm_{\p\gamma}\csm^{-1}_{\gamma\gamma}\csm_{\gamma\p}}\,.
\end{equation}
Using \eref{eq.3.1} and \eref{eq.3.1inv}, the second determinant gives
\begin{equation}
\det\paren{\csm_{\p\p}-\csm_{\p\gamma}\csm^{-1}_{\gamma\gamma}\csm_{\gamma\p}}=
-\frac{\sig\gamma^{\sfrac{1}{2}}}{N  s }.\label{det1}
\end{equation}
The remaining determinant $\det\left(\csm_{\gamma\gamma}\right)$ can be
expressed in terms of the determinant of the DeWitt metric, 
\begin{eqnarray}
\det\left(\csm_{\gamma\gamma}\right)&=\det\paren{-\frac{\sig U}{2 N}\gamma^{1/2}G^{abcd}}\nonumber\\
&=\paren{-\frac{\sig U}{2 N}}^{\frac{d(d+1)}{2}}\det\left(\gamma^{1/2}G^{abcd}\right)\label{det2},
\end{eqnarray}
where we have used that the space of symmetric rank two tensor fields is $d(d+1)/2$ dimensional.
The trace of the unit matrix in configuration space gives
\begin{equation}
\text{tr}\,\tensor*{\delta}{^A_B}=\frac{d(d+1)}{2}+1\,.
\end{equation}
The determinant of the DeWitt metric $\gamma^{1/2}G^{abcd}$ is well known \cite{DeWitt:1967yk}, and can be obtained by variation of the DeWitt metric $\gamma^{1/2}G^{abcd}$ with respect to $\gamma_{ab}$. The result reads
\begin{equation}
\det\left(\gamma^{1/2}G^{abcd}\right)=-\alpha\left(\gamma^{1/2}\right)^{\frac{(d+1)(d-4)}{2}}\label{det3}\,.
\end{equation}
Here $\alpha$ is some positive constant \cite{DeWitt:1967yk}. 
The explicit value of this constant is irrelevant for our purposes as it cancels in all relevant expressions.
Combining \eref{det1} with \eref{det2} and \eref{det3}, we obtain for the determinant of the configuration space metric (for $d\geq3$)
\begin{equation}
\csm
=\frac{\alpha\sig}{N s }\left(\frac{\gamma^{1/2}\,U}{2N}\right)^{\frac{d(d+1)}{2}}\left(\gamma^{1/2}\right)^{-(2d+1)}\,,\label{detcsm}
\end{equation}
were we have used $\sig^2=1$ in the last equality.
The determinant  $\csm$ can also be obtained by different methods. As a consistency check, we calculate $\csm$ from the metric \eref{eq.3.1} and the Christoffel symbols \eref{ccsm1}-\eref{ccsm6} 
as the solution to the differential equation
\begin{equation}
\delta_{A}\ln\csm^{1/2}=\Gamma^{B}_{AB}\,.
\end{equation}  
We find the same expression \eref{det2}. In this approach $\alpha$ arises as constant of integration. 
Note that in the purely gravitational case, the signature of the configuration space metric (DeWitt metric) is indefinite, independent of the signature of spacetime. In contrast, \eref{detcsm} implies that the signature of the configuration space metric $\csm_{AB}$ does depend on the signature $\sig$ of the metric $g_{\mu\nu}$ in the $D$-dimensional ambient space ${\cal M}$,
 due to the extra scalar degree of freedom $\p$ in configuration space.

\subsection{Riemann tensor configuration space}
Given the expressions for the Christoffel symbols \eref{ccsm1}-\eref{ccsm6}, 
it is straightforward to calculate the non-vanishing components of the configuration space Riemann tensor 
\begin{equation}
\tensor{\csriemann}{^A_{BCD}}
=\delta_C\cschrist^A_{BD}-\delta_D\cschrist^A_{BC}+\cschrist^A_{CE}\cschrist^E_{BD}-\cschrist^A_{DE}\cschrist^E_{BC}. 
\end{equation}
It is convenient to express the components of the Riemann tensor with two indices raised ${\tensor{\csriemann}{^A^B_C_D}=\tensor{\csm}{^{BF}}\tensor{\csriemann}{^A_F_C_D}}$. We find four independent non-vanishing components
\begin{eqnarray}
\fl\tensor{\csriemann}{_{abcd}^{efgh}}
=-\frac{\sig N\gamma^{-\sfrac{1}{2}}}{4(d-1)}
\left[\frac{d}{U}-\frac{2}{d-1} s \left(\frac{ U_1}{U}\right)^2\right]\delta_{[[ab}^{[[ef}\delta_{cd]]}^{gh]]}\nonumber\\
+\frac{\sig N\gamma^{-\sfrac{1}{2}}}{2(d-1)}
\left[\frac{1}{U}-\frac{2}{d-1} s \left(\frac{U_{1}}{U}\right)^2\right]\delta_{[[ab}^{[[ef}\gamma^{\phantom{\dagger}}_{cd]]}
\gamma_{\phantom{\dagger}}^{gh]]}\nonumber\\
+\frac{2\sig\gamma^{-\sfrac{1}{2}}}{U}
\delta_{(a}^{[[(e}\gamma_{\phantom{\dagger}}^{f)(g}\delta_{(c}^{h)]]}\gamma^{\phantom{\dagger}}_{d)b)}\,,\\
\fl\tensor{\csriemann}{_{abcd}^{ef\p}}
=\frac{2\sig N\gamma^{-\sfrac{1}{2}}}{(d-1)^2} s \frac{U_{1}}{U}\left[\frac{d-1}{4 s  U}
+\frac{ s_1 }{2 s }\frac{U_{1}}{U}
-\left(\frac{U_{1}}{U}\right)^2
+\frac{U_{2}}{U}\right]\delta_{[[ab}^{ef}\gamma^{\phantom{\dagger}}_{cd]]},\\
\fl\tensor{\csriemann}{_{ab}^{\p efgh}}
=\frac{\sig N\gamma^{-\sfrac{1}{2}}}{4(d-1)} s  \frac{U_{1}}{U}
\delta_{ab}^{[[ef}\gamma_{\phantom{\dagger}}^{gh]]}\,,\\
\fl\tensor{\csriemann}{_{ab}^{\p ef}_\p}
=-\frac{\sig N\gamma^{-\sfrac{1}{2}}}{2(d-1)} s \left[\frac{d}{4 s  U}+\frac{ s _{1}}{2 s }\frac{U_1}{U}-\frac{2d-1}{2(d-1)}\left(\frac{U_1}{U}\right)^2+\frac{U_2}{U}\right]\delta_{ab}^{ef}\nonumber\\
 +\frac{\sig N\gamma^{-\sfrac{1}{2}}}{4(d-1)} s \left[\frac{1}{2 s  U}-\frac{1}{d-1}\left(\frac{U_1}{U}\right)^2\right]\gamma_{ab}\gamma^{ef}\,.
\end{eqnarray}
Note, that we have introduced the compact notation for the (anti)symmetrization of a symmetric index pair
\begin{eqnarray}
\fl A_{((ab}B_{cd))}:=&\frac{1}{2}\left(A_{ab}B_{cd}+A_{cd}B_{ab}\right),\quad A_{[[ab}B_{cd]]}:=&\frac{1}{2}\left(A_{ab}B_{cd}-A_{cd}B_{ab}\right)\,.
\end{eqnarray}

\subsection{Ricci tensor configuration space}
The configuration Ricci tensor can then be found by contracting the first and third
indices
\begin{eqnarray}
\tensor{\csricci}{^A_B}=\tensor{\csriemann}{^{CA}_{CB}}.
\end{eqnarray}
In components, we obtain
\begin{eqnarray}
\fl \tensor{\csricci}{_{ab}^{cd}}
=-\frac{\sig N\gamma^{-\sfrac{1}{2}} s }{2 (d-1)}
   \left[\frac{d(d^2-1)}{8 s  U}+\frac{ s_1 }{2 s }\frac{U_1}{U}-\frac{(d+4)}{4}\left(\frac{U_1}{U}\right)^2+\frac{U_2}{U}\right]\delta_{ab}^{cd}\nonumber\\
+\frac{\sig N\gamma^{-\sfrac{1}{2}} s }{8}\left[\frac{d-6}{2 s  U}-\frac{d+2}{d-1}\left(\frac{U_1}{U}\right)^2\right]\gamma_{ab}\gamma^{cd},\\
\fl \tensor{\csricci}{_{ab}_\p}
=\frac{\sig(d+2)N\gamma^{-\sfrac{1}{2}} s }{2(d-1)}\frac{U_{1}}{U}
\left[\frac{d-1}{4 s  U}+\frac{ s_1 }{2 s }\frac{U_{1}}{U}
-\left(\frac{U_{1}}{U}\right)^2+\frac{U_{2}}{U}\right]\gamma_{ab}\,,\\
\fl \tensor{\csricci}{^\p^a^b}
=\frac{\sig(d+2)N\gamma^{-\sfrac{1}{2}} s }{16}\frac{U_{1}}{U}\gamma^{ab}\,,\\
\fl \tensor{\csricci}{^\p_\p}
=-\frac{\sig dN\gamma^{-\sfrac{1}{2}} s }{4}
   \left[\frac{d+2}{4 s  U}+\frac{d+1}{2(d-1)}\frac{ s_1 }{ s }\frac{U_1}{U}\right.
\left.-\frac{2d+3}{2(d-1)}\left(\frac{U_1}{U}\right)^2+\frac{d+1}{d-1}\frac{U_2}{U}\right].
\end{eqnarray}

\subsection{Ricci scalar configuration space}
The configuration space Ricci scalar is obtained by contracting the indices of the configuration space Ricci tensor ${\csricci=\tensor{\csricci}{^{A}_{A}}}$,
\begin{eqnarray}
\fl
\csricci=-\sig \frac{d(d+2)(d^2-7d+8)}{32}\frac{N\gamma^{-\sfrac{1}{2}}}{U}\nonumber\\
-\sig \frac{d(d+1)}{4(d-1)}N\gamma^{-\sfrac{1}{2}} s \left[\frac{ s_1 }{ s }\frac{U_{1}}{U}-\frac{(d+6)}{4}\left(\frac{U_{1}}{U}\right)^2+2\frac{U_{2}}{U}\right].\label{csriccis}
\end{eqnarray}
The purely gravitational contribution to the Ricci scalar with $U=U_0$ is given by
\begin{equation}
\csricci_{\mathrm{grav}}=-\sig N\gamma^{-\sfrac{1}{2}}\,\frac{d(d+2)(d^2-7d+4)}{32\,U_0}.
\end{equation}
For $\sig=-1$, $d=3$, $N=1$ and $U_0=1/2$, we can compare our result for $\csricci_{\mathrm{grav}}$ with the result obtained \cite{DeWitt:1967yk}. We obtain
\begin{equation}
\csricci_{\mathrm{grav}}=-\gamma^{-\sfrac{1}{2}}\frac{15}{4}\,.\label{CSRicciSgrav}
\end{equation}
This disagrees with \cite{DeWitt:1967yk}, where it was found in that $\csricci_{\mathrm{grav}}$ is three times the value of \eref{CSRicciSgrav}.

\section{Conformal transformations}\label{App:ConformalTransformation}
The conformal transformation from a metric $g_{\mu\nu}$ to a new metric $\tilde{g}_{\mu\nu}$ on a $D$-dimensional manifold is given by
\begin{equation}
g_{\mu\nu}=\Omega\,\tilde{g}_{\mu\nu}\,.\label{CTmetric}
\end{equation}
Here, $\Omega(\p)$ is the conformal factor, which is a strictly positive function of the scalar field $\p$.
The transformation \eref{CTmetric} implies
\begin{equation}
g^{\mu\nu}=\Omega^{-1}\,\tilde{g}^{\mu\nu},\qquad g^{1/2}=\Omega^{D/2}\,\tilde{g}^{1/2},\qquad \christ^\rho_{\mu\nu}=\tilde{\christ}^\rho_{\mu\nu}+\tensor{\Omega}{^\rho_{\mu\nu}},
\end{equation}
where $\christ^\rho_{\mu\nu}$ are the Christoffel symbols of the metric $g$, $\tilde{\christ}^\rho_{\mu\nu}$ the Christoffel symbols of the metric $\tilde{g}_{\mu\nu}$ and $\tensor{\Omega}{^\rho_{\mu\nu}}$ the difference tensor, which includes derivatives of the conformal factor,  
\begin{equation}
\tensor{\Omega}{^\rho_{\mu\nu}}:=
\frac{1}{2\Omega}\left(\tensor*{\delta}{_\mu^\rho}\partial^{\phantom{a}}_\nu\Omega
+\tensor*{\delta}{_\nu^\rho}\partial^{\phantom{a}}_\mu\Omega
-\tensor{\tilde{g}}{_{\mu\nu}}\tensor{\tilde{g}}{^{\rho\alpha}}\partial_\alpha\Omega\right).
\end{equation}
The Riemann tensor is found to be
\begin{eqnarray}
\tensor{\riemann}{^\rho_{\sigma\mu\nu}}={}&\tensor{\tilde{\riemann}}{^{\rho}_{\sigma\mu\nu}}
+\Omega^{-1}\left(\tensor{\tilde{g}}{^{\rho\alpha}}\tensor{\tilde{g}}{_{\sigma[\mu}}\tensor*{\delta}{_{\nu]}^\beta}-\tensor*{\delta}{_{[\mu}^\rho}\tensor*{\delta}{_{\nu]}^\alpha}\tensor*{\delta}{_\sigma^\beta}\right)\tensor{\tilde{\nabla}}{_\alpha}\tensor{\tilde{\nabla}}{_\beta}\Omega\nonumber\\
&+\frac{\Omega^{-2}}{2}\left(3\tensor*{\delta}{_{[\mu}^\rho}\tensor*{\delta}{_{\nu]}^\alpha}\tensor*{\delta}{_\sigma^\beta}-3\tensor{\tilde{g}}{_{\sigma[\mu}}\tensor*{\delta}{_{\nu]}^\alpha}\tensor{\tilde{g}}{^{\rho\beta}}+\tensor*{\delta}{_{[\mu}^\rho}\tensor{\tilde{g}}{_{\nu]\sigma}}\tensor{\tilde{g}}{^{\alpha\beta}}\right)\tensor{\tilde{\nabla}}{_\alpha}\Omega\tensor{\tilde{\nabla}}{_\beta}\Omega\,.\label{CTRiemann}
\end{eqnarray}
The Ricci tensor is obtained from \eref{CTRiemann} by contracting the first and third indices
\begin{eqnarray}
\ricci_{\mu\nu}
={}&\tensor{\tilde{\ricci}}{_{\mu\nu}}-\frac{\Omega^{-1}}{2}\left[(D-2)\tensor*{\delta}{_\mu^\alpha}\tensor*{\delta}{_\nu^\beta}
+\tensor{\tilde{g}}{_{\mu\nu}}\tensor{\tilde{g}}{^{\alpha\beta}}\right]
\tensor{\tilde{\nabla}}{_\alpha}\tensor{\tilde{\nabla}}{_\beta}\Omega\nonumber\\
&+
\frac{\Omega^{-2}}{4}\left[3(D-2)\tensor*{\delta}{_\mu^\alpha}\tensor*{\delta}{_\nu^\beta}
-(D-4)\tensor{\tilde{g}}{_{\mu\nu}}\tensor{\tilde{g}}{^{\alpha\beta}}\right]
\tensor{\tilde{\nabla}}{_\alpha}\Omega
\tensor{\tilde{\nabla}}{_\beta}\Omega\,.\label{CTRicciT}
\end{eqnarray}
Finally, the Ricci scalar is found from \eref{CTRicciT} by contracting the Ricci tnesor with the inverse metric $g^{\mu\nu}$,
\begin{eqnarray}
\ricci={}&\Omega^{-1}\tilde{\ricci}
-(D-1)\Omega^{-2}\tensor{\tilde{g}}{^{\alpha\beta}}\tensor{\tilde{\nabla}}{_\alpha}\tensor{\tilde{\nabla}}{_\beta}\Omega\nonumber\\
&-\frac{(D-1)(D-6)}{4}\Omega^{-3}\tilde{g}{^{\alpha\beta}}
\tensor{\tilde{\nabla}}{_\alpha}\Omega\tensor{\tilde{\nabla}}{_\beta}\Omega\,.\label{CTRicciS}
\end{eqnarray}

\section{Transformation to the Einstein frame}\label{App:JFtoEF}
The action \eref{eq.2.1}, expressed in terms of the fields $(g_{\mu\nu},\varphi)$ features a non-minimal coupling to gravity. It can be brought into a form which resembles that of a scalar field minimally coupled to Einstein gravity, by a nonlinear field redefinition $(g,\p)\to(\tilde{g},\tilde{\p})$.
We first perform a conformal transformation \eref{CTmetric} of the metric field $g_{\mu\nu}\to\tilde{g}_{\mu\nu}$ with the conformal factor 
\begin{equation}
\Omega=\left(\frac{U}{U_0}\right)^{-\frac{2}{D-2}}\,.\label{ConfFac}
\end{equation}
The Ricci scalar transforms according to \eref{CTRicciS}.
Inserting the transformation for the metric into the JF action \eref{eq.2.1}, we find
\begin{eqnarray}
S[\tilde{g},\varphi]&=\int\rmd^Dx\sqrt{\sig\tilde{g}}\left(U_0\tilde{R}-\frac{1}{2}\frac{U_0}{U}\frac{GU+2\frac{D-1}{D-2}U_{1}^2}{U}\tensor{\tilde{\nabla}}{_\mu}\p\tensor{\tilde{\nabla}}{^\mu}\p-\tilde{V}\right),
\end{eqnarray}
where we have defined the EF potential
\begin{equation}
\tilde{V}:=\left(\frac{U}{U_0}\right)^{-\frac{D}{D-2}} V\,.\label{EFPot}
\end{equation}
It is always understood that indices of EF tensors, which carry a tilde, are raised or lowered with respect to the metric $\tilde{g}_{\alpha\beta}$ and that covariant derivatives $\tilde{\nabla}_{\mu}$ are defined with respect to the Christoffel symbol of the metric $\tilde{g}_{\alpha\beta}$.
The kinetic term can be brought into canonical form by a redefinition $\varphi\to\tilde{\varphi}$ of the JF scalar field $\varphi$ to the EF scalar field $\tilde{\varphi}$, defined by
\begin{equation}
\left(\frac{\partial\tilde{\p}}{\partial\p}\right)^2=\frac{U_0}{U}\frac{GU+\frac{2d}{d-1}U_{1}^2}{U}=\left(\frac{U s }{U_0}\right)^{-1}\,.\label{EFScalField}
\end{equation}
We have used the definition \eref{eq.3.3} of the $s$-factor in the last equality.
In terms of the EF parametrization $(\tilde{g},\tilde{\varphi})$, the action \eref{eq.2.1} reads
\begin{equation}\label{einstein frame}
S[\tilde{g},\tilde{\varphi}]=\int\,\rmd^Dx\sqrt{\sig\tilde{g}}\left(\frac{M_{\mathrm{P}}^2}{2}\tilde{R}-\frac{1}{2}\tensor{\tilde{\nabla}}{_\mu}\tilde{\p}\tensor{\tilde{\nabla}}{^\mu}\tilde{\p}-\tilde{V}\right)\,.
\end{equation}
In the last step, we have identified the constant $U_0=M_{\mathrm{P}}^2/2$ with the Planck mass.

\section*{References}

\bibliography{WDW_V2}{}
\bibliographystyle{iopart-num}

\end{document}